\documentclass[ALICE,manyauthors]{cernphprep}
 \pdfoutput=1
\usepackage[comma,square,numbers,sort&compress]{natbib}
\usepackage{hyperref}
\usepackage{lineno}
\usepackage{xspace}
\usepackage{lineno}
\usepackage{hyperref}
\usepackage{multirow}

\usepackage{textcomp}
\usepackage{amsmath}  
\usepackage{xcolor}   
\usepackage{listings}
\usepackage{rotating}
\usepackage{placeins}
\usepackage{algpseudocode}
\usepackage{graphicx}
\usepackage[T1]{fontenc}

\usepackage{natbib}
\usepackage[toc,page]{appendix}

\newcommand{\Nsame}{\ensuremath{N_{\mathrm{same}}}}
\newcommand{\Nmixed}{\ensuremath{N_{\mathrm{mixed}}}}

\newcommand{\ptgamma}{\ensuremath{p_{\mathrm{T}}^\gamma}}

\newcommand{\zt}{\ensuremath{z_{\mathrm{T}}}\xspace}

\newcommand{\sqrts}{\ensuremath{\sqrt{s}}}

\newcommand{\lambdasquare}{\ensuremath{\sigma^{2}_{\mathrm{long}}}}

\newcommand{\ydecay}{\ensuremath{\gamma^\mathrm{decay}}}
\newcommand{\gammaiso}{\ensuremath{\gamma^\mathrm{iso}}}

\newcommand{\sqrtsNN}{\ensuremath{\sqrt{s_\mathrm{NN}}}}

\newcommand{\pPb}          {\mbox{p--Pb}\xspace}

\newcommand{\snn}          {\ensuremath{\sqrt{s_{\mathrm{NN}}}}\xspace}
\newcommand{\pt}           {\ensuremath{p_{\rm T}}\xspace}

\newcommand{\nineH}        {$\sqrt{s}~=~0.9$~Te\kern-.1emV\xspace}
\newcommand{\seven}        {$\sqrt{s}~=~7$~Te\kern-.1emV\xspace}
\newcommand{\twoH}         {$\sqrt{s}~=~0.2$~Te\kern-.1emV\xspace}
\newcommand{\twosevensix}  {$\sqrt{s}~=~2.76$~Te\kern-.1emV\xspace}
\newcommand{\five}         {$\sqrt{s}~=~5.02$~Te\kern-.1emV\xspace}
\newcommand{\twosevensixnn}{$\sqrt{s_{\mathrm{NN}}}~=~2.76$~Te\kern-.1emV\xspace}
\newcommand{\fivenn}       {$\sqrt{s_{\mathrm{NN}}}~=~5.02$~Te\kern-.1emV\xspace}

\newcommand{\GeVc}         {Ge\kern-.1emV/$c$\xspace}
\newcommand{\MeVc}         {Me\kern-.1emV/$c$\xspace}
\newcommand{\TeV}          {Te\kern-.1emV\xspace}
\newcommand{\GeV}          {Ge\kern-.1emV\xspace}
\newcommand{\MeV}          {Me\kern-.1emV\xspace}
\newcommand{\GeVmass}      {Ge\kern-.2emV/$c^2$\xspace}
\newcommand{\MeVmass}      {Me\kern-.2emV/$c^2$\xspace}

\begin{document}
\begin{titlepage}
\PHyear{2020}       
\PHnumber{097}      
\PHdate{29 May}  

\title{Measurement of isolated photon--hadron correlations in \snn = 5.02 TeV pp and \pPb~collisions}
\ShortTitle{Isolated photon--hadron correlations in 5.02 TeV pp and \pPb collisions}   

\Collaboration{ALICE Collaboration\thanks{See Appendix~\ref{app:collab} for the list of collaboration members}}
\ShortAuthor{ALICE Collaboration} 

\begin{abstract}
This paper presents isolated photon--hadron correlations using
pp and \pPb data collected by the ALICE detector at the LHC. For photons with $|\eta|<0.67$ and $12 < \pt < 40$ \GeVc, the associated yield of charged particles in the range {$|\eta|<0.80$} and $0.5 < \pt < 10$ \GeVc is presented. These momenta are much lower than previous measurements at the LHC. No significant difference between pp and~\pPb~is observed, with \textsc{Pythia 8.2} describing both data sets within uncertainties. This measurement constrains nuclear effects on the parton fragmentation in~\pPb~collisions, and provides a benchmark for future studies of Pb--Pb collisions.

\end{abstract}
\end{titlepage}

\setcounter{page}{2} 

\section{Introduction}
\label{sec:introduction}
Understanding the dynamics of quarks and gluons in nucleons and nuclei is a key goal of modern nuclear physics. Proton$-$nucleus (pA) collisions at high energies provide information about the parton structure of nuclei, parton$-$nucleus interactions, and parton fragmentation in a nuclear medium~\cite{Accardi:2009qv}. The energy of the Large Hadron Collider (LHC) available for pA collisions is a factor of 25 larger than at the Relativistic Heavy Ion Collider (RHIC), and thus it provides unprecedented reach in longitudinal momentum fraction Bjorken-$x$ and $Q^{2}$~\cite{Salgado:2011wc}. 

Parton fragmentation may be modified in the nucleus, offering a way to explore the dynamics of QCD in nuclei including elastic, inelastic, and coherent multiple scattering of partons. Moreover, the known spatial dimensions of nuclei provide a filter possibly shedding light on the timescale of the fragmentation process, which remains unknown~\cite{Accardi:2009qv,Accardi:2012qut}. Additionally, because photons produced in hard scatterings do not strongly interact, they constrain the parton kinematics from the same scattering before any modification. Thus, measurements of photon-tagged jet fragmentation in pA collisions serve as a powerful tool to study multiple-scattering effects in cold nuclear matter~\cite{Xing:2012ii}, which serve as a control for effects of the quark$-$gluon plasma (QGP) in nucleus$-$nucleus collisions, where modifications of the jet spectrum, fragmentation, and substructure have been observed~\cite{Connors:2017ptx}.

Traditionally, the effects attributed to the QGP 
were expected to be absent in pA collisions. However, recent measurements show evidence for collective behavior~\cite{Nagle:2018nvi}, which might hint that a small droplet of QGP forms in pA collisions, yet no significant modification of jet production or fragmentation has been found.   


In di-hadron and direct photon-hadron correlations, no significant modification of the jet fragmentation was observed in measurements by the PHENIX collaboration in d--Au collisions at a center-of-mass energy of 200 GeV~\cite{Adler:2005ad} and the ALICE collaboration in \pPb~collisions at 5.02 TeV~\cite{Acharya:2018edi,Adam:2015xea} at mid rapidity. At forward rapidity, a strong-modification was observed by the PHENIX collaboration in d-Au collisions~\cite{Adare:2011sc}. A recent measurement by the PHENIX collaboration with pp, p--Al, and p--Au data revealed a transverse momentum broadening consistent with a path-length dependent effect~\cite{Aidala:2018eqn}. However, a recent ATLAS measurement of the jet fragmentation function in \pPb~collisions showed no evidence for modification of jet fragmentation for jets with $45<\pt<206$ \GeVc~\cite{Aaboud:2017tke}.
Measurements of the fragmentation of jets with much lower momentum are necessary to limit the Lorentz boost to the timescales of fragmentation, as such a boost may result in fragmentation outside the nucleus. These measurements would test the $Q^{2}$ evolution of fragmentation functions in cold nuclear matter, testing factorization theorems that are neither proven nor expected to hold in general for collisions involving nuclei~\cite{deFlorian:2011fp}. 


In this work, azimuthal correlations of charged hadrons with isolated photons, $\gammaiso$, are analyzed in \pPb~and pp collisions with a center-of-mass energy of \sqrtsNN~ = 5.02 TeV. Isolated photons are measured at midrapidity, {$|\eta|<0.67$}, and with transverse momenta in the range $12 <\pt<40$ \GeVc, which yields the scaling variable {$x_{\mathrm{T}} = 2\pt/\sqrt{s_{\mathrm{NN}}}\ = $ 0.005--0.016}. The kinematic range probed in this analysis offers access to a lower $Q^{2}$ than other LHC experiments, which is where the largest nuclear effects can be expected,  and to a similar $x_{\mathrm{T}}$ range as RHIC measurements at forward rapidity~\cite{Adare:2011sc}. 

The measurement of the transverse momentum of $\gammaiso$ constrains the recoiling parton kinematics in a way that is not possible with inclusive jet production and provides an effective way to probe the nuclear modification of the fragmentation function. Moreover, the per-trigger yield is the ratio of a semi-inclusive cross-section (photon + jet) and inclusive cross-section (photon). Both quantities are sensitive to the nuclear parton distribution functions (PDF) in the same way \cite{kang2016semiinclusive,gamma_cross_section}. Thus, by measuring per-photon quantities, sensitivity to the nuclear PDF is eliminated.


%



This paper is organized as follows: Section~\ref{sec:experimentalsetup} covers the experimental setup; the datasets and simulations are presented in Section~\ref{sec:datasets}; isolated photon and charged hadron reconstructions are detailed in Sections \ref{sec:tracking} and \ref{sec:photon}; the purity measurement is reported in Section \ref{sec:purity}; Section~\ref{sec:correlations} describes the correlation measurements; Section~\ref{sec:systematics} reports the systematic uncertainties of the measurement; Section~\ref{sec:results} presents the results; and the conclusions are discussed in Section~\ref{sec:conclusions}.
\section{Experimental setup}
\label{sec:experimentalsetup}
A comprehensive description of the ALICE experiment and its performance is provided in Refs.~\cite{Aamodt:2008zz,Abelev:2014ffa}. The detector elements most relevant for this study are the electromagnetic calorimeter system, which is used to measure and trigger on high $\pt$ photons, and the inner tracking system, which is used for tracking and determination of the interaction vertex. Both are located inside a large solenoidal magnet with a field strength of 0.5 T along the beam direction. 

The Electromagnetic Calorimeter (EMCal) is a sampling calorimeter composed of 77 alternating layers of {1.4 mm} lead and {1.7 mm} polystyrene scintillators. 
 It has a cellular structure made up of square cells with a transverse size of 6 $\times$ 6 cm$^{2}$. 
Wavelength shifting fibers attached to the perpendicular faces of each cell collect the scintillation light. These fibers are then connected to Avalanche Photodiodes (APDs) which amplify the generated scintillation light.

The EMCal is located at a radial distance of approximately 428 cm from the nominal interaction point, and its cell granularity is $\Delta\eta\times\Delta\varphi$ = 14.3$\times$14.3 mrad. Its energy resolution is $\sigma_{E}/E = A\oplus B/\sqrt{E} \oplus C/E$ where \textit{A} = 1.7\%, \textit{B} = 11.3\%, \textit{C} = 4.8\%, and the energy $E$ is given in units of GeV~\cite{Abeysekara:2010ze}. The linearity of the response of the detector and electronics has been measured with electron test beams to a precision 
of a few percent for the momentum range probed in this analysis. The non-linearity is negligible for cluster energies between 3 and 50 GeV, which is the relevant range for this analysis. The geometrical acceptance of the EMCal is $|\eta|<0.7$ and  $80^{\circ} < \varphi < 187^{\circ}$.

The Di-jet Calorimeter (DCal) is an extension of the EMCal. It is back-to-back in azimuth with respect to the EMCal and uses the same technology and material as the EMCal \cite{Allen:2010stl}. Thus, it has identical granularity and intrinsic energy resolution. It covers $0.22< |\eta|<0.7$ and $260^{\circ}< \varphi <320^{\circ}$, and an additional region between $|\eta|<$0.7 and $320^{\circ}<\varphi<327^{\circ}$. It was installed and commissioned during the first long shutdown of the LHC and therefore was operational during the 2017 pp run but not during the 2013 \pPb~run. Thus, both the EMCal and the DCal are used in the trigger and analysis of the pp collisions, while only the EMCal was used in \pPb.


The inner tracking system (ITS) consists of six layers of silicon detectors and is located directly around the interaction point. 
The two innermost layers consist of silicon pixel detectors positioned at radial distances of 3.9 cm and 7.6 cm, followed by two layers of silicon drift detectors at 15.0 cm and 23.9 cm, and two layers of silicon strip detectors at 38.0 cm and 43.0 cm. The ITS covers $|\eta|<0.9$ and has full azimuthal coverage. 

The V0 detector is used to provide the minimum bias trigger and to estimate the particle multiplicity in each event. The detector consists of two scintillator arrays, V0A and V0C, located on opposite sides of the interaction point at $z=+340$ cm and $z=-90$ cm and covering $2.8 <\eta < 5.1$ and $-3.7 <\eta < -1.7$, respectively. 
\section{Datasets}
\label{sec:datasets}
The data used for this analysis were collected during the 2013 \pPb run and the 2017 pp run, both at a center-of-mass energy of $\sqrt{s_{\mathrm{NN}}}=5.02$ TeV. 
Photon events were selected by the ALICE EMCal trigger. The EMCal issues triggers at two different levels, Level 0 (L0) and Level 1 (L1). The events that pass L0 selection are further processed at L1. The L0 decision, issued at most 1.2 $\mu$s after the collision, is based on the analog charge sum of 4 $\times$ 4 adjacent cells evaluated with a sliding window algorithm within each physical Trigger Region Unit (TRU) spanning 8 $\times$ 48 cells in coincidence with a minimum bias trigger. The L1 trigger decision, which must be taken within 6.2 $\mu$s after the collision, can incorporate additional information from different TRUs, as well as other triggers or detectors. Additionally, the L1 extends the 4$\times$4 sliding window search across neighboring TRUs, resulting in a roughly 30\% larger trigger area than the L0 trigger \cite{Acharya_2017}. In 2013 p-Pb collisions, one L0 and two L1 triggers with different thresholds were used. The L0 threshold was 3 GeV, while the L1 thresholds were 11~GeV and 7~GeV. In pp collisions, an L0 threshold of 2.5 GeV and a single L1 threshold of 4 GeV were used. This analysis requires clusters with an energy above 12 GeV in order to avoid the usage of the triggers around their respective threshold values in pp and \pPb.



Due to the 2-in-1 magnet design of the LHC, which requires the same magnetic rigidity for both colliding beams, the beams had different energies per nucleon. The energy of the protons was 4 TeV. In the lead nucleus, the energy per nucleon was {$1.56$ TeV $= (Z/A) \times$ 4 TeV}, where $Z=82$ is the atomic number of lead and $A =$ 208 is the nuclear mass number of the lead isotope used. This energy asymmetry results in a rapidity boost of the nucleon$-$nucleon center-of-mass frame by 0.465 units relative to the ALICE rest frame in the direction of the proton beam. 

Full detector simulations are used in the study of the tracking performance described in Section~\ref{sec:tracking}, in the purity measurement with template fits described in Section~\ref{sec:purity}, and for comparisons with data described in Section~\ref{sec:results}. The simulations of hard processes are based on the \textsc{PYTHIA} 8.2 event generator, 2013 Monash Tune~\cite{Sjostrand:2007gs}. In \textsc{PYTHIA}, the signal events are included via $2\to2$ matrix elements with $gq\to\gamma q$ and $q\overline{q}\to\gamma g$ hard scatterings, defined at the leading order, followed by the leading-logarithm approximation of the parton shower and hadronization. To simulate \pPb~events, the pp dijet and gamma-jet events simulated with PYTHIA 8.2 are embedded into \pPb~inelastic collision events generated with \textsc{DPMJET}~\cite{Roesler:2000he} to reproduce the experimentally measured global \pPb~event properties. The simulated data include only those events with a calorimeter cluster above threshold, and are boosted by 0.465 units of rapidity in the nucleon-nucleon center-of-mass frame.

The detector response is simulated with \textsc{GEANT3}~\cite{Brun:1994aa} where the generated events are processed through the same reconstruction chain as the data. Following Ref.~\cite{Acharya:2019jkx}, a correction is applied to the \textsc{GEANT} simulation to mimic the observed cross-talk between calorimeter cells, which is attributed to the readout electronics. This correction leads to a good description of the electromagnetic showers observed in data.

To ensure a uniform acceptance and reconstruction efficiency in the pseudorapidity region $|\eta| < 0.8$, only events with a reconstructed vertex within $\pm10$ cm of the center of the detector along the beam direction are used. 
\section{Tracking performance}
\label{sec:tracking}
The data taking approach during part of the 2017 pp run was to read out only a subset of the ALICE detector systems. This enhanced the sampled luminosity by reading out at a higher rate. This lightweight readout approach included the EMCal and the ITS but excluded the Time Projection Chamber. As a result, ITS-only tracking is used for both pp and \pPb~data in this measurement. This approach differs from the standard ALICE tracking, but it has also been used for dedicated analyses of 
low momentum particles that do not reach the TPC~\cite{Aamodt:2011zj}. Previous  studies using standalone ITS tracking used a maximum track \pt\ of 0.8 \GeVc \cite{Contin_2012}. What is novel in this analysis is the use of an extended range of \pt\ in the ITS-only tracking from 0.5 to 
10 \GeVc. 

All tracks are required to fulfill the following conditions: at least 4 hits in the ITS detector, a distance of closest approach to the primary vertex in the transverse plane 
{less than 2.4 cm},  a distance of closest approach along the beam axis {less than 3.2 cm}, and a track fit quality cut for ITS track points which satisfy $\chi^{2}_\mathrm{ITS}/N^\mathrm{hits}_\mathrm{ITS} < 36$.
 
Monte Carlo simulations are used to determine the efficiency and purity for primary charged particles \cite{ALICE-PUBLIC-2017-005}.
In \pPb~collisions, the tracking efficiency is 87$\%$ for tracks with 1 $< \pt <$~10 \GeVc, decreasing to roughly 85$\%$ at $\pt$ = 0.5 GeV/c; the momentum resolution is 6.6$\%$ for $\pt$ = 0.5~\GeVc and 13$\%$ for $\pt$~=~10~\GeVc. In pp collisions, the tracking efficiency is 85$\%$ for tracks at 1 $< \pt <$ 10 \GeVc decreasing to roughly 83$\%$ at $\pt$ = 0.5~\GeVc, with a momentum resolution of 6.6$\%$ for \pt~= 0.5 \GeVc~and 15$\%$ for \pt= 10 \GeVc. The fake track rate in \pPb~is 1.9\% at 0.5 \GeVc, growing linearly with \pt,  reaching 19\%  at 10 \GeVc. For tracks in pp, the fake rate is 2.6\% at 0.5 \GeVc~and grows linearly to 18\% at 10~\GeVc. 



The following check on the simulation was performed to ensure that it reproduces minimum$-$bias data. As the yield of charged particles in minimum$-$bias data is generally independent of $\varphi$, any dips in the $\varphi$ distribution are clearly visible in both simulation and data. After efficiency corrections, the $\varphi$ distribution is flat within $\pm$~2.5$\%$. $\varphi$ and $\eta$ detector-dependent effects on the cluster-track pair acceptance are corrected with the event mixing technique described in Section \ref{sec:correlations}.

To validate the combined effect of tracking efficiency, fake rate, and track momentum smearing corrections obtained from simulation of ITS-only tracking, the published charged-particle spectrum in \pPb~collisions at {$\sqrt{s_{\mathrm{NN}}}=$ 5.02 TeV} from Ref.~\cite{Acharya:2018qsh} was reproduced. The published spectrum was obtained using the ALICE standard tracking and is compatible with ITS-only tracking within $\pm$8\% for $\pt<0.85$ \GeVc and $\pm$5\% for $0.85<\pt<10$ \GeVc. This difference is taken into account in the systematic uncertainty assigned to tracking corrections. 

\section{Isolated photon selection}
\label{sec:photon}
The signal for this analysis is isolated prompt photons. At the lowest order in pQCD, prompt photons are produced via two processes: (i) quark-gluon Compton scattering, $qg \to q\gamma$, (ii) quark-antiquark annihilation, $q\overline{q} \to g\gamma$, and, with a much smaller contribution,  $q\overline{q}\to$ $\gamma\gamma$. In addition, prompt photons are produced by higher-order processes, such as fragmentation or bremsstrahlung \cite{AURENCHE199334}. The collinear part of such processes has been shown to contribute effectively also at lowest order.



\subsection{Isolation requirement}
At leading order in pQCD, prompt photons are produced in 2$\to$2 processes surrounded by very little hadronic activity, while fragmentation photons are found within a jet. Beyond leading order, the direct and fragmentation components 
cannot be factorized; the sum of their cross sections is the physical observable. However, theoretical calculations can be simplified through the use of an isolation requirement \cite{PhysRevD.82.014015}, which also helps to suppress the background from decays of neutral mesons often found within jets.

The isolation variable for this analysis is defined as the scalar sum of the transverse momentum of charged particles within an angular radius, $R =\sqrt{(\Delta\varphi)^{2} +(\Delta\eta)^{2}  } =0.4$, around the cluster direction. In contrast with a previous ALICE isolated photon measurement, Ref.~\cite{Acharya:2019jkx}, the isolation variable does not include neutral particles. This enables us to use the full acceptance of the EMCal and reduces biases arising from correlation with the opening angle of $\pi^{0}$ decays. However, it does result in a slightly lower purity of the isolated single photon signal. 

For the determination of the isolation criterium,~$\pt^\mathrm{iso}$, the background due to the underlying event is estimated with the Voronoi method from the \textsc{FastJet} jet area/median package~\cite{Cacciari:2009dp} on an event-by-event basis and subtracted according to:
\begin{equation}
\pt^\mathrm{iso} = \sum_{\mathrm{track}~\in\Delta R<0.4} p_{\mathrm{T}}^{\mathrm{track}} - \rho \times \pi\ \times 0.4^{2}.
\label{eq:isoraw}
\end{equation}

The charged-particle density, $\rho$, is calculated for each event; average values are 3.2 \GeVc in photon- triggered events in \pPb and 1.6 \GeVc in pp collisions. 
 A requirement of $\pt^\mathrm{iso}<1.5$ \GeVc is used, which results in a signal efficiency of about 90$\%$ that does not significantly depend on the photon $\pt$. For photons near the edge of the detector, the isolation energy requirement is scaled to account for any missing area in the isolation cone. Given that the results presented in this analysis are normalized to the number of reconstructed photons, the $\gammaiso$ efficiency does not affect the measurement. Effects from $\varphi$ and $\eta$ dependence of the tracking performance on the isolation cut were found to be negligible. 

\subsection{Cluster selection}
The photon reconstruction closely follows the method described in Ref.~\cite{Acharya:2019jkx}. Clusters are obtained by grouping all adjacent cells with common sides whose energy is above {100 MeV}, starting from a seed cell with at least {500 MeV}. Furthermore, a cluster must contain at least two cells to remove single-cell electronic noise fluctuations. Clusters are required to have a minimum \pt of \ptgamma $\geq 12$ \GeVc. The time of the highest-energy cell in the clusters relative to the main bunch crossing must satisfy $\Delta t < 20$ ns to reduce out-of-bunch pileup. In order to limit spurious signals caused by particles hitting the EMCal APDs, clusters are required to have $E_{\mathrm{cross}}/E_{\mathrm{cluster}}>0.05$, where $E_{\mathrm{cross}}$ is the sum of the energy in the cells adjacent to, but not including, the the leading cell, and $E_{\mathrm{cluster}}$ is the total energy of the entire cluster. The number of local maxima in the cluster is required to be less than three to reduce hadronic background.

Clusters originating from isolated, prompt photons are separated from background arising from neutral meson decays by means of the distinct shape of the electromagnetic shower that is encoded in the \lambdasquare variable, which represents the extent of the cluster. The \lambdasquare~variable is defined as the square of the larger eigenvalue of the energy distribution in the $\eta$--$\varphi$ plane:
\begin{equation}
\lambdasquare = (\sigma^{2}_{\varphi\varphi} + \sigma^{2}_{\eta\eta})/2 + \sqrt{(\sigma^{2}_{\varphi\varphi} - \sigma^{2}_{\eta\eta})/4 + \sigma^{2}_{\varphi\eta}},
\end{equation}

where $\sigma^{2}_{\varphi \eta} = \langle \varphi\eta \rangle - \langle \varphi \rangle\langle \eta \rangle$ are the covariance matrix elements; the integers $\varphi,\eta$ are cell indices along the $\hat{\eta}$ and  $\hat{\varphi}$ axes; $\langle \varphi\eta \rangle$ and $\langle \varphi\rangle$, $\langle \eta\rangle$ are the second and the first moments of the cluster position cell.  The position is weighted by $\mathrm{max}\left(\log(E_{\mathrm{cell}}/E_{\mathrm{cluster}}) - w_{0},0\right).$ Following previous work~\cite{Acharya:2018dqe}, the cutoff in the log-weighting is chosen to be $w_{0}=-4.5$. Cells that contain less than {$e^{-4.5} =$ 1.1$\%$} of the total cluster energy are not considered in the $\lambdasquare$ calculation. Thus, $\lambdasquare$ discriminates between clusters belonging to single photons, having a $\lambdasquare$ distribution which is narrow and symmetric, and merged photons from neutral meson decays, which are asymmetric and have a distribution dominated by a long tail towards higher values. 

Most single-photon clusters yield $\lambdasquare\approx 0.25$, as shown in Figure \ref{fig:TemplateFit} where the signal is displayed in blue and the background is displayed in yellow. Figure 1 is discussed in more detail in section 6.  Consequently, a cluster selection of $\lambdasquare<0.30$ is applied irrespective of \pt. Simulations indicate this results in a signal efficiency of about 90$\%$ with no significant \pt~dependence.

The main background 
remaining after the cluster and isolation cuts arises from multijet events where one jet typically contains a $\pi^{0}$ or $\eta$ that carries most of the jet energy and the decay photons are misidentified as single photons. The magnitude of this background is quantified in Section \ref{sec:purity}.

\section{Purity measurement}
\label{sec:purity}

\begin{figure*}
    \includegraphics[width=0.32\textwidth]{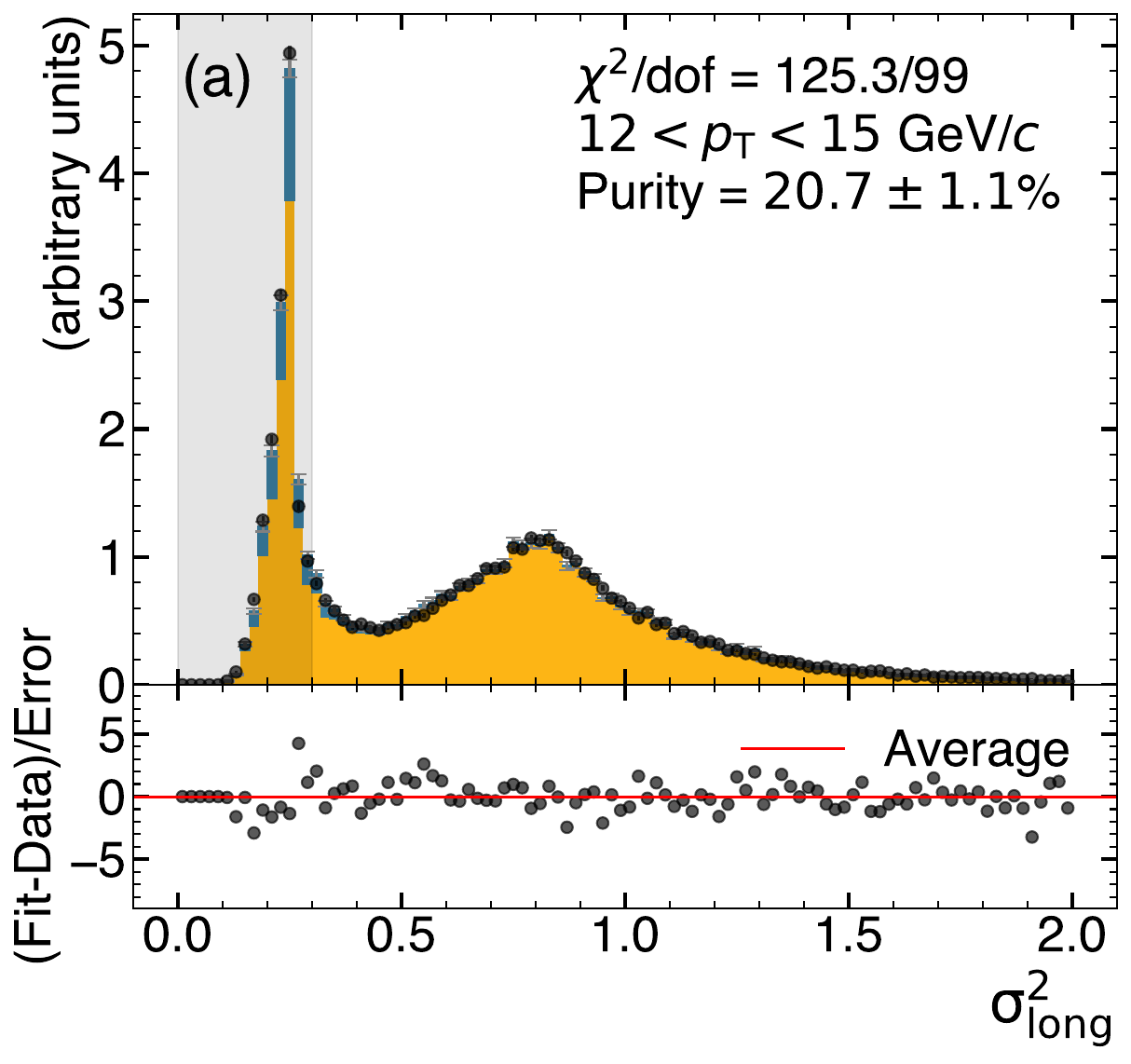}
        \includegraphics[width=0.32\textwidth]{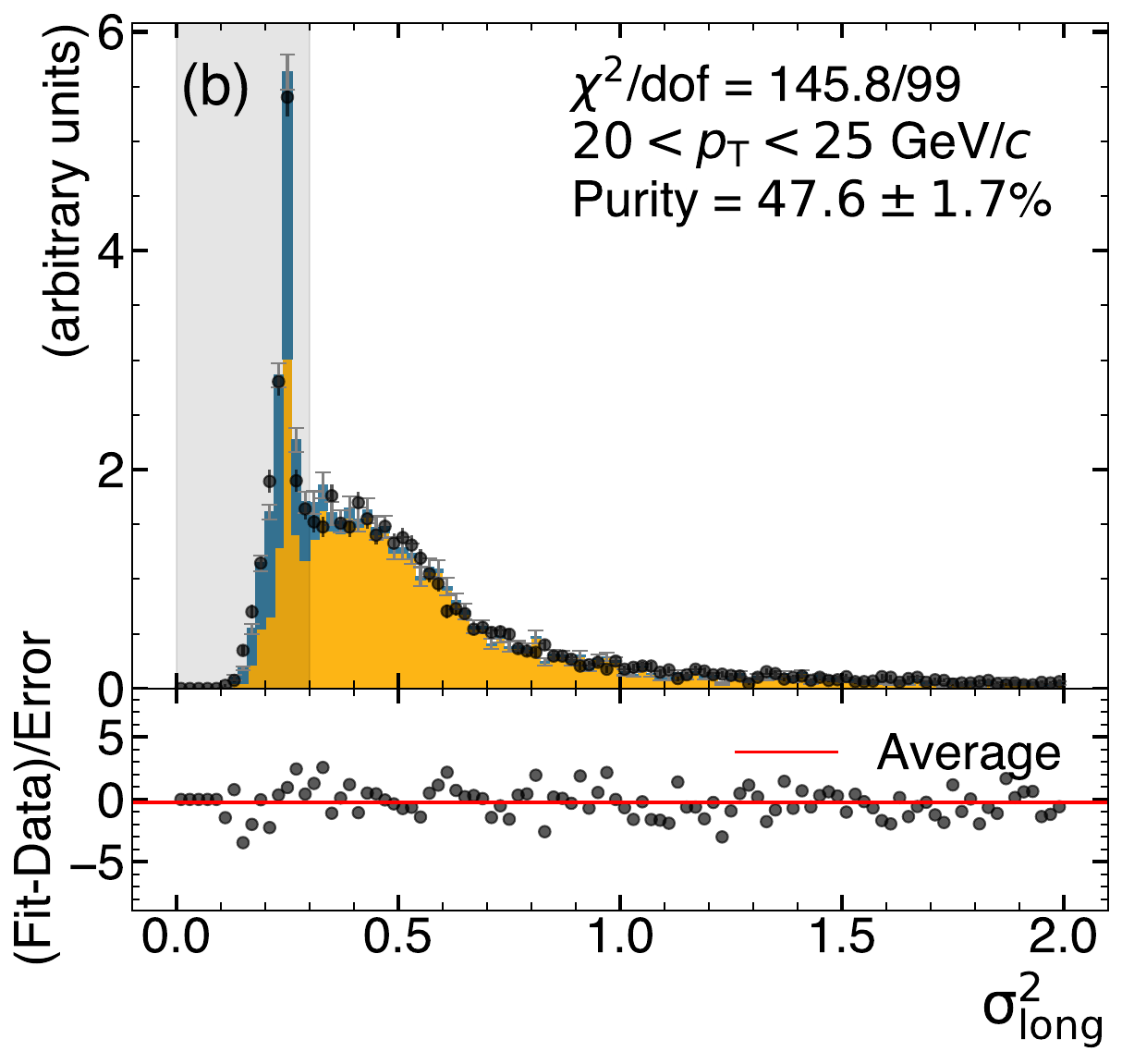}
    \includegraphics[width=0.32\textwidth]{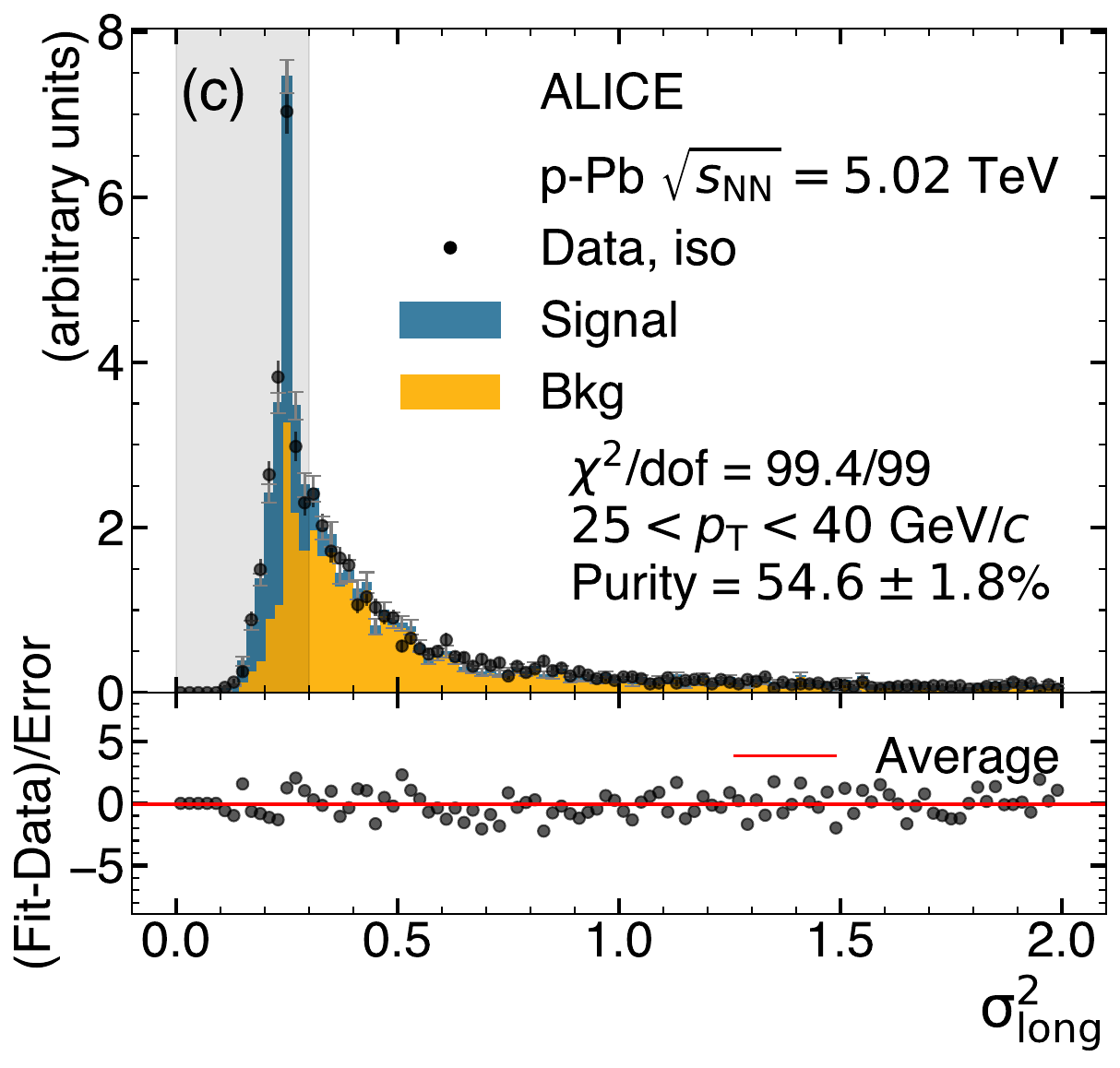}
    \caption{\lambdasquare distribution of isolated clusters (black) and template fit results for \pPb~data in various \pt~ranges. The stacked histograms (yellow for background, blue for signal) show the predicted counts corresponding to the best fit. The bottom panels show the normalized residuals of the fit, with the statistical uncertainty on the isolated cluster data and the background template added in quadrature. The gray shaded region indicates the signal region for the isolated-photon selection. See text for additional details.}
    \label{fig:TemplateFit}
\end{figure*}

The purity of the $\gammaiso$ candidate sample is measured using a two-component template fit. The $\lambdasquare$ distribution for the isolated cluster sample is fit with a linear combination of the signal distribution, determined from a photon-jet simulation, and the background distribution, determined from data using an anti-isolated sideband ($5.0 < \pt^\mathrm{iso} < 10.0$ \GeVc) and corrected using a dijet simulation.

The \textsc{MINUIT}~\cite{James:1975dr} package is used for $\chi^{2}$ minimization and the \textsc{MIGRAD} package for uncertainty estimation. The only free parameter in the fit is the number of signal clusters, $N_{\mathrm{sig}}$, because the overall normalization, $N$, is fixed to the total number of isolated clusters:
\begin{equation}
N^{\mathrm{observed}}(\lambdasquare) = N_{\mathrm{sig}}\times S(\lambdasquare) + (N-N_{\mathrm{sig}})\times B(\lambdasquare),
\end{equation}
where $S(\lambdasquare)$ and $B(\lambdasquare)$ are the normalized signal and background templates. Examples of template fits are shown in Figure~\ref{fig:TemplateFit}.
The peaks observed in the background templates originate mostly from collinear or very asymmetric $\pi^{0}\to\gamma\gamma$ decays. Photons from $\eta$ decays also contribute to the peaks in the background template.




The background template is corrected for a bias due to correlations between the shower-shape and isolation variables \cite{Khachatryan:2010fm}. 
This correlation leads to clusters in the isolation sideband having a somewhat higher hadronic activity than the true isolated background. Consequently, a background template constructed from this sideband region has an increased number of background-like clusters and
purity values obtained using this
 systematically overestimate the true purity.
A correction for this bias, $R(\lambdasquare)$, is determined using dijet simulated events which also contain the correlation between trigger photon shower-shape and isolation cut.
The ratio of the shower-shape distributions of clusters in the signal (Iso, $\pt^\mathrm{iso} < 1.5$ \GeVc) region and sideband (Anti-iso, $5.0 < \pt^\mathrm{iso} < 10.0$ \GeVc) region is constructed via


\begin{equation}
    R(\lambdasquare)=\frac{\text{Iso}_{\text{MC}}(\lambdasquare)}{\text{Anti-iso}_{\text{MC}}(\lambdasquare).}
    \label{eq:bkgtemplateweights}
\end{equation}
This ratio of shower shape distributions is applied as a multiplicative correction to the background template:

\begin{equation}
    B^{\text{corr.}}(\lambdasquare)=\text{Anti-iso}_{\text{data}}(\lambdasquare)\times R(\lambdasquare).
    \label{eq:bkgtemplatecorrection}
\end{equation}

\begin{figure}
    \centering
    \includegraphics[width=0.55\textwidth]{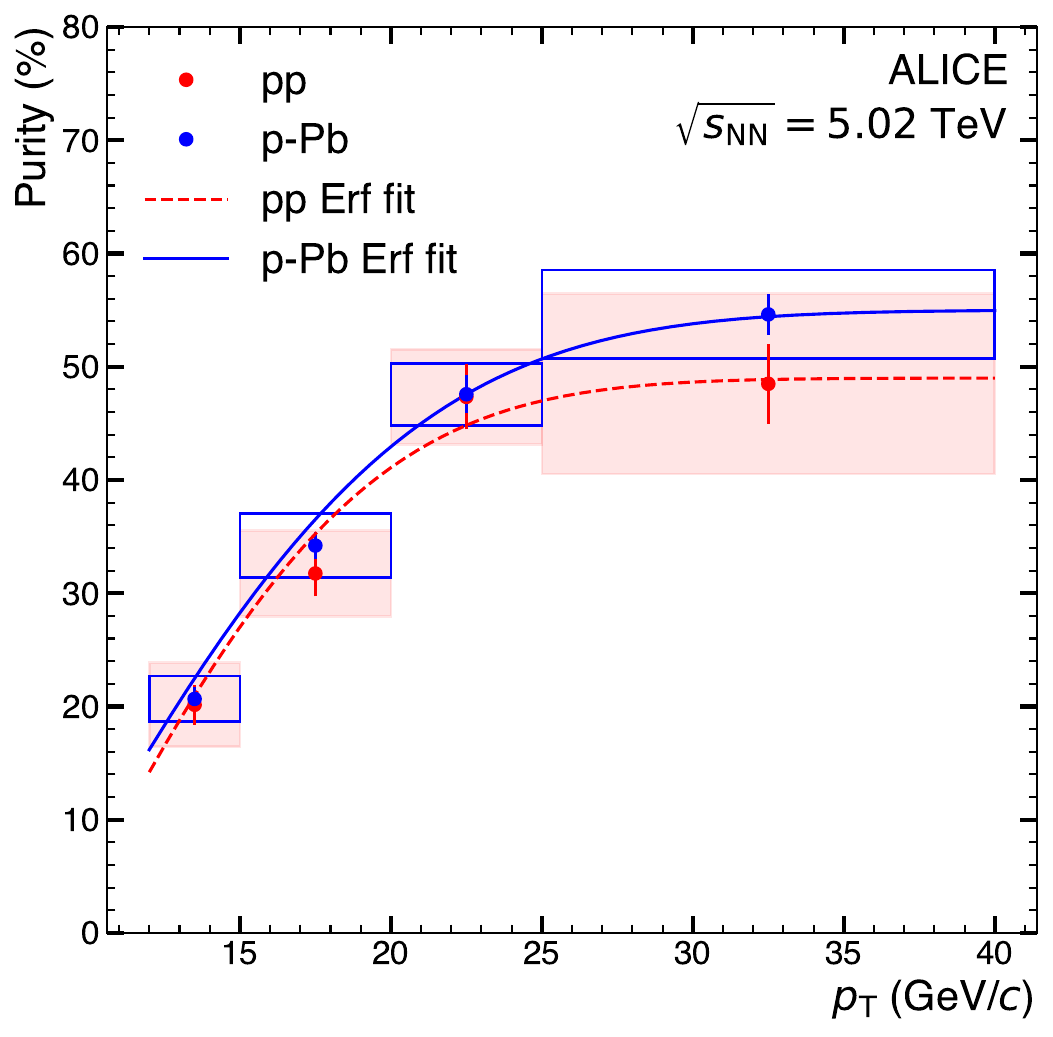}
    \caption{Purity of the $\gammaiso$ sample as a function of transverse momentum for pp (red) and \pPb~(blue) data. The error bars represent statistical uncertainties only. The red shaded area represents systematic uncertainties in pp, while the blue empty boxes represent systematic uncertainties in \pPb. The smooth lines correspond to a three-parameter error function fit to the data.}
    \label{fig:Purity}
\end{figure}

This background template correction results in an absolute correction on the purity of 8$\%$--14$\%$ depending on the cluster $\pt$. The purities as a function of the cluster $\pt$ are shown in Figure~\ref{fig:Purity}. They are compatible between the pp and \pPb~datasets within the uncertainties. A three-parameter error function is fit to the data. The fits have been checked with several bin variations to ensure that they accurately represent the quickly rising purity at low \pt.

\section{Azimuthal Correlations}
\label{sec:correlations}
The analysis of the correlation functions proceeds as follows:
the angular correlation of $\gammaiso$ candidates with charged particles is  constructed, requiring photons within {$|\eta|<0.67$} and {$12 < \pt < 40$} \GeVc and associated charged particles within {$|\eta|<0.80$} and {$0.5 < \pt < 10$} \GeVc. 
Geometrical acceptance effects are corrected using a mixed-event correlation function, as described in detail below. The contribution of {$\ydecay$--hadron} correlations 
is subtracted using the {$\ydecay$--hadron} correlation function determined by inverting the cluster shower-shape selection to select clusters with large values of \lambdasquare. The {$\ydecay$--hadron} correlation is scaled and subtracted from the isolated photon-hadron correlation function.
Next, the remaining contribution from the underlying event is subtracted. This uncorrelated background is estimated using the zero-yield-at-minimum (ZYAM) method \cite{Adler_2006}, where a background pedestal is estimated from the minimum of the correlation function.  The ZYAM background level is cross-checked using a control region at large $|\eta^{\mathrm{h}}-\eta^{\gamma}|$.
The away-side of each fully subtracted and corrected correlation function is then integrated to measure the conditional yield of away-side hadrons.
This analysis is performed in intervals of $\zt \equiv \pt^\mathrm{h}/\ptgamma$ for charged particles, such that the measurement of the away-side yield is sensitive to the parton fragmentation function.

Event mixing is used as a data-driven approach to correct for detector acceptance effects. By constructing observables with particles from different events, true physics correlations are removed from the correlation functions, leaving only the detector effects resulting from limited acceptance in \(\eta\) and detector inhomogeneities in $\eta$ and $\varphi$. Events are classified in bins of multiplicity (V0 amplitude, sum of V0A and V0C signals) and primary vertex $z$-position. Typically, event mixing uses event pairs within these bins. In this analysis, however, events are paired that are on-average closer in multiplicity and $z$-position than the standard binning method. This is accomplished using the Gale-Shapley stable matching algorithm \cite{GALE1985223} that removes the need for binning. The same-event correlation function in each $\zt$ bin is then divided by the corresponding mixed-event correlation function. 

The pair-acceptance corrected correlation function is given by:


\begin{equation}
\label{eq:Y}
C(\Delta \varphi, \Delta \eta) = \frac{S(\Delta \varphi, \Delta \eta)}{M(\Delta \varphi, \Delta \eta)},
\end{equation}

where $S(\Delta \varphi, \Delta \eta)$ is the same-event correlation, and $M(\Delta \varphi, \Delta \eta)$ is the mixed-event correlation. $S(\Delta \varphi, \Delta \eta)$ is calculated by: 
\begin{equation}
S(\Delta \varphi, \Delta \eta) = \frac{1}{N_{\mathrm{\gammaiso}}}\frac{\mathrm{d}^2N_{\mathrm{same}}(\Delta \varphi, \Delta \eta)}{\mathrm{d}\Delta \varphi \mathrm{d}\Delta \eta},
\end{equation}

with $N_{\gammaiso}$~as the number of clusters that pass the isolation and shower shape cuts, and \Nsame~ as the number of same event cluster-track pairs. $\mathrm{d}^2\Nsame/\mathrm{d}\Delta \varphi \mathrm{d}\Delta \eta$ is found by pairing trigger particles with tracks from the same event. The mixed-event distribution, $M(\Delta \varphi, \Delta \eta)$, is given by 
\begin{equation}
M(\Delta \varphi, \Delta \eta) = \alpha \frac{\mathrm{d}^2 \Nmixed(\Delta \varphi, \Delta \eta)}{\mathrm{d}\Delta \varphi \mathrm{d}\Delta \eta},
\end{equation}

where $\alpha$ is the normalization constant that sets the maximum value of the mixed event correlation to unity, and \Nmixed~is the number of mixed event cluster-track pairs. The term $\mathrm{d}^2 \Nmixed/\mathrm{d}\Delta \varphi \mathrm{d}\Delta \eta$ is obtained by pairing trigger particles from \(\gamma\)-triggered events with tracks from minimum bias events matched in $z$-vertex and multiplicity. The number of events was chosen such that any uncertainty from event mixing is negligible.

The tracks used in the same-event correlation functions, $S(\Delta \varphi, \Delta \eta)$, are corrected for single track acceptance, efficiency, and $\pt^{\mathrm{track}}$ bin-to-bin migration 
calculated from the simulations. The corrections are implemented using track-by-track weighting when filling the correlation histograms. The weights are given by:

\begin{equation}
	w_{\mathrm{tracking}}(\pt^{\mathrm{track}}) = \frac{1}{\epsilon}\times(1- f)\times b,
	\label{eq:track_weights}
\end{equation}

where $\epsilon$ is the track efficiency and $f$ is the fake rate. $b$ is the bin-to-bin migration factor that corrects for \pt\ smearing arising from the finite $\pt^\mathrm{track}$ resolution and is determined by taking the ratio of the reconstructed \pt\ and the true \pt\ for all true tracks as a function of $\pt^\mathrm{true}.$
The efficiency, fake rate, and bin migration corrections are applied in bins of $\pt^\mathrm{track}$.

\begin{figure*}
    \centering
    \includegraphics[width=0.5\textwidth]{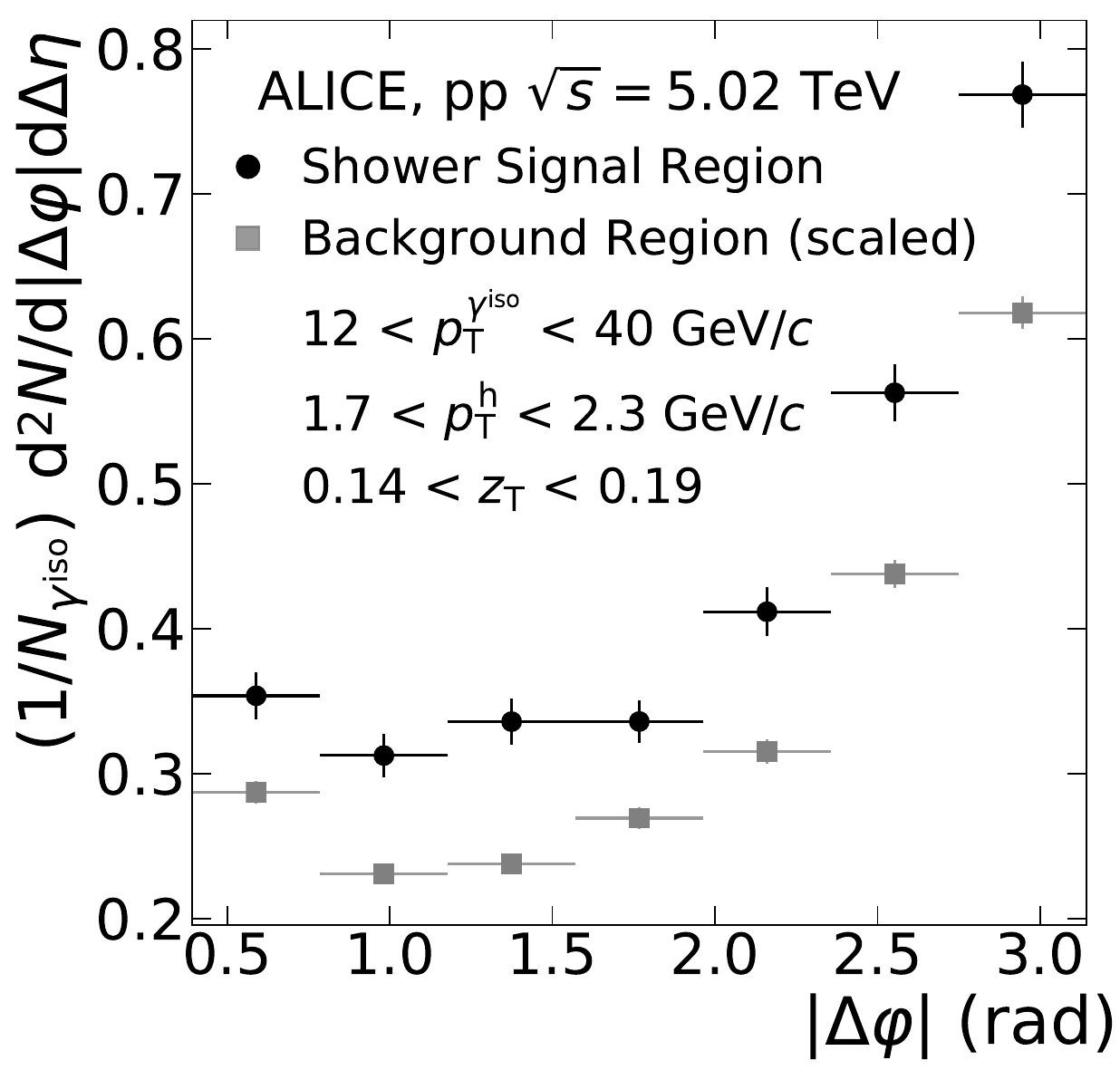}
        
    \caption{$\gammaiso$--hadron signal region (black circles) and background region (grey squares) correlations in pp collisions at \sqrts~= 5.02 TeV as measured by the ALICE detector. The shower signal region photons correspond to isolated clusters with \lambdasquare $< 0.3$, while the shower background region photons correspond to isolated clusters with \lambdasquare $> 0.4$. The vertical bars represent statistical uncertainty only. The horizontal bars represent the bin width in $|\Delta\varphi|$. The background correlation is subtracted from the signal correlation according to the numerator in Eq. \ref{Corr_Subtraction}.}
    \label{fig:SR_BR_Overlay}
\end{figure*}

After this correction,
the contribution to the signal region correlation function from decay photons that pass the cluster selection is subtracted. The shower signal region photons correspond to isolated clusters with $\sigma^2_\mathrm{long} < 0.3$. The subtraction of the correlated background starts by inverting the shower shape criteria ($\sigma^2_\mathrm{long} > 0.4$) to select isolated clusters that arise primarily from neutral meson decays. The correlation of these shower background region clusters and associated hadrons is measured ($C_\mathrm{BR}$). This $\ydecay$--hadron correlation function is scaled by ($1-\mathrm{Purity}$) and subtracted from the shower signal region correlation function ($C_\mathrm{SR}$) according to: 

\begin{equation}
\label{Corr_Subtraction}
C_\mathrm{S} = \frac{C_\mathrm{SR}-(1-P)C_\mathrm{BR}}{P},    
\end{equation}
\FloatBarrier

where $P$ is the purity and $C_\mathrm{S}$ is the signal correlation function we aim to measure. $(1-P)C_\mathrm{BR}$ corresponds to the contribution of decay photons to the signal region correlation function after isolation and shower shape cuts.
The quantities $C_\mathrm{SR}$ and $(1-P)C_\mathrm{BR}$ are shown in Fig.~\ref{fig:SR_BR_Overlay}.
The overall factor of 1/$P$ in Eq.~\ref{Corr_Subtraction} is used to obtain the correct per-trigger yields after the $\ydecay$--hadron contribution has been subtracted. The scaling of the correlations is done cluster-by-cluster, with the shower signal and shower background region clusters scaled by 1/$P$ and $\frac{1-P}{P}$, respectively, according to Eq.~\ref{Corr_Subtraction}. The purity used in the cluster-by-cluster weighing procedure is determined by fitting the purity values from Fig.~\ref{fig:Purity} to a three-parameter error function in order to avoid bin-edge effects and capture the quickly-rising behavior of the purity at low cluster $\pt$.

To ensure that the shower background region correlations properly estimate the decay photons within the shower signal region, the background region cluster $\pt$ distribution is weighted to match the signal region cluster $\pt$ distribution. This has no significant effect on the background subtraction, indicating that the background shape varies slowly with $\pt$ and discrepancies between $\pt$ distributions for background and signal triggers have no significant effect on the correlations.

The uncorrelated background from the underlying event is estimated in two ways. In the ZYAM procedure, the average of the correlation function in the range $ 0.4 < |\Delta\varphi| <\frac{\pi}{2}$ is taken as the uncorrelated background estimate. This range takes advantage of the fact that there is no near-side jet peak in isolated photon-hadron correlations. As a result, the correlation function for $|\Delta\varphi| < \frac{\pi}{2}$ should contain minimal signal. The correlation function for $|\Delta\varphi| < 0.4$ is not used for the underlying event estimation to avoid any bias from the isolation region.
The second method to estimate the underlying event takes the average value of the correlation function in the range $0.8 < \Delta\eta < 1.4$ and $0.4 < |\Delta\varphi| < 1.2$. Both methods yield background estimates compatible within statistical uncertainties. The ZYAM method is used in the final pedestal subtraction due to the method's smaller statistical uncertainty.
\section{Systematic uncertainties}
\label{sec:systematics}
 The following sources of systematic uncertainty in the \gammaiso--hadron measurement have been considered: uncertainty on the purity measurement, underlying event subtraction, ITS-only tracking performance, acceptance mismatch due to the boost in \pPb~ relative to pp, the \gammaiso~\pt~spectra, and the photon energy scale. The systematic uncertainties in the \gammaiso--hadron and fragmentation measurements are described in more detail in this section and are summarized in Table \ref{tab:BigSummarySystematics}.

 \begin{table*}
  \centering
  \caption{Summary of uncertainties in $\gammaiso$-hadron correlations, which are reported as per-trigger yields of correlated hadrons. 
The ranges shown encompass the relative uncertainties for hadron \zt~in two ranges: Low-\zt ($0.06<\zt<0.18$) and High-\zt ($0.18<\zt<0.6$). The statistical uncertainty in the underlying event estimate using the ZYAM method is shown here. Uncertainties arising from the detector material budget, luminosity scale, vertex efficiency, trigger corrections, and photon reconstruction do not contribute to the final uncertainty.} 
  \begin{tabular*}{1.0\textwidth}{@{\extracolsep{\fill}}lcccc@{}}
    \hline
     & pp (Low-\zt) & pp (High-\zt) & \pPb(Low-\zt) &\pPb (High-\zt) \\
  \hline
  Statistical Uncertainty & 19--40\% & 28--49\% & 16--23\% & 27--44\% \\
  \hline 
  Photon Purity & 18\% & 18\% & 11\% & 11\% \\
  Underlying Event & 8\%--15\% & 7\%--12\% & 7\%--9\% & 8\%--9\% \\
  Tracking performance &  5.6\% & 5.6\% &  5.6\% & 5.6\% \\
  Acceptance mismatch &-- & -- &2\% & 2\% \\ 
  Photon Energy Scale & $<$1\% & $<$1\%  & $<$1\% & $<$1\%\\
  Photon Energy Resolution & $<$1\% & $<$1\%  & $<$1\% & $<$1\%\\
  Material budget & $<$1\% & $<$1\% & $<$1\% & $<$1\% \\
  \hline
  Total Systematic Uncertainty: & 21\%--24\% & 20\%--22\% & 14\%--16\% & 15\%\\
  \hline
  Total Uncertainty & 28\%--47\% & 34\%--54\% & 22\%--27\% & 31\%--46\% \\
  \hline
  \end{tabular*}
  \label{tab:BigSummarySystematics}
\end{table*}
\FloatBarrier
\subsection{Purity}
The three sources of systematic uncertainty on the purity are the background template correction, construction of the signal template, and the choice of the anti-isolation region. These sources of systematic uncertainty on the purity measurement are summarized in Table~\ref{tab:pursyst}. No single source of uncertainty dominates across \pt~ranges or collision systems. These are summed in quadrature to get an absolute overall systematic uncertainty on the purity of 2--8\%.

To estimate the uncertainty on the background template correction, the ratio in Eq. \ref{eq:bkgtemplateweights} is also constructed in data and combined to create a double ratio:

\begin{equation}
    \text{Double ratio} = \frac{\text{Iso}_{\text{data}}/\text{Anti-iso}_{\text{data}}}{\text{Iso}_{\text{MC}}/\text{Anti-iso}_{\text{MC}}}.
    \label{eq:bkgtemplatedoubleratio}
\end{equation}

In the signal region of the shower shape distribution ($0.1<\lambdasquare<0.3$), this double ratio will be far from unity, as the data have prompt photons and the dijet MC do not. However, away from that region, where the background dominates, the double ratio should be flat in \lambdasquare~if the dijet MC reproduces the background shower-shape of the data. A linear function is fit to this double ratio in the background-dominated region of the shower shape distribution. The linear function is then extrapolated back into the signal region. To estimate the systematic uncertainty on the background template correction, that linear fit and its variation within its fit uncertainty are used as additional multiplicative factors in Eq.~\ref{eq:bkgtemplateweights}. The purities calculated with these modified background template corrections are used to estimate the systematic uncertainty on the purity from the background template correction.

To estimate the uncertainty on the signal template, a background-only template fit is performed and compared to the full template fit. For the background-only fit, the background template is fit to the data in the background-dominated region of the shower shape distribution. This fixes the normalization of the background template. Then,  in the signal region, the difference between the data and background is used to calculate the purity, with no contribution from the signal template. The difference between this purity and the purity as calculated with the signal template is taken to be the uncertainty on the signal template.

To estimate the uncertainty from the anti-isolation selection, a template fit is performed with background templates built from different overlapping anti-isolation selections. This identifies a nominal anti-isolation sideband selection where the template fits are good and the purities are stable. The uncertainty is estimated from the spread of the purities calculated from the template fits for which the anti-isolation selection falls within the nominal anti-isolation selection ($5 < \pt^\mathrm{iso} < 10$ \GeVc).

\begin{table}
\caption{Summary of the purity and its systematic uncertainties (absolute quantities) on the $\gammaiso$ selection. The range spans the uncertainties on the purity in different \ptgamma~ bins.}
   \begin{tabular*}{1.0\columnwidth}{@{\extracolsep{\fill}}lcc@{}}
    \hline
        & pp  & \pPb\\
        Purity &20-49\% & 21-53\%\\
        \hline
        Background template correction & 2.9--3.4\% & 1.2--2.1\% \\
        Signal distribution & 0.8--5.9\% & 1.1--2.3\% \\
        Anti-isolation selection &  1.2--4.0\% & 0.8--2.4\% \\
        \hline
        Total &  3.7--7.9\% & 2.0--3.9\%
    \end{tabular*}
    \label{tab:pursyst}
\end{table}
The uncertainty in the purity measurement is propagated to the correlation function measurement following Eq.~\ref{Corr_Subtraction}. The resulting uncertainty on the correlation function is $\pm18\%$ for pp data and  $\pm11\%$ for \pPb~data. A large fraction of the total uncertainty in the purity is either statistical uncertainty or systematic uncertainties that arise due to limited data sample. Therefore, uncertainties arising from the purity in the pp and \pPb~data are largely uncorrelated in the $\gamma$-hadron analysis. To be conservative, they are taken to be totally uncorrelated. The uncertainty on the purity in pp is larger than in p--Pb due to the pp dataset having lower statistics: the background templates are directly obtained from data, and the uncertainty on the signal template is evaluated using data as well.

\subsection{Underlying Event Subtraction}
The uncertainty in the underlying event subtraction originates from statistical fluctuations in the ZYAM estimate and propagates directly to the per-trigger hadron yields. This uncertainty ranges from 7\% to 15\% depending on the \zt~bin and data set. The uncertainty is fully correlated in $\Delta\varphi$ for a given \zt~bin, but totally uncorrelated among \zt~bins. It is also uncorrelated between the pp and \pPb~datasets.

\subsection{Track reconstruction}
The uncertainty due to charged-particle $\pt^\mathrm{track}$ reconstruction determined by comparing the stand-alone ITS $\pt^\mathrm{track}$ specta with published ALICE $\pt^\mathrm{track}$~spectra using standard ITS+TPC tracking \cite{Acharya:2018qsh}.
As described in Section~\ref{sec:tracking}, the combined uncertainty due to tracking efficiency, fake rate, and bin-to-bin migration corrections amounts to $\pm5\%$ added in quadrature with the total systematic uncertainty of the reference $\pt$ spectra. This systematic effect in the reference $\pt$ spectra is 1.6\%$-$1.9\% in pp collisions, and 2.1\%$-$2.5\% in \pPb collisions, for tracks with $0.5<\pt^\mathrm{track}<10$ \GeVc\cite{Acharya:2018qsh}. 

Systematic uncertainties due to secondary-particle contamination and from modeling of the particle composition in Monte Carlo simulations are small ($<2\%$) for the range $0.5<\pt<10$ \GeVc. These were already estimated in Ref. \cite{Acharya:2018qsh} for the pp and \pPb~datasets and are already included in the reference spectrum systematic uncertainty estimate described above. The tracking performances in the pp and \pPb~datasets are very similar, but as a conservative approach these systematic uncertainties are treated as completely uncorrelated.

\subsection{Rapidity Boost}
The difference between the energy of the proton and the energy of the nucleons in the Pb nucleus yields a boost of the center-of-mass of $\Delta y = 0.465$ in the proton-going direction. This means that in \pPb~collisions, the acceptance for photons of $-0.67<\eta<0.67$ corresponds to $-0.2<\eta<1.14$ in the center-of-mass frame, whereas the charged-particle acceptance of $-0.8<\eta<0.8$ corresponds to $-0.33<\eta<1.27$ in the center-of-mass frame.
\textsc{Pythia8} events are used to generate \gammaiso--hadron correlations for isolated photons within $-0.20<\eta<1.14$  and charged particles within $-0.33<\eta<1.27$. This is then compared to \gammaiso--hadron correlations using the nominal ranges of $-0.67<\eta<0.67$ and $-0.8<\eta<0.8$ for isolated photons and charged particles, respectively. These studies of \gammaiso--hadron correlations show that the impact of an acceptance mismatch between pp and \pPb~data is about $5\%$, independent of $\zt$. This estimate is subject to PDF uncertainties, which dictate the shape of the differential cross section in pseudorapidity of photons and associated hadrons. A correction is applied for this effect and an additional 2$\%$ systematic uncertainty on the per-trigger hadron yields is assigned. This systematic uncertainty is taken to be completely correlated with \zt~and is assigned only to the \pPb~measurements. 

\subsection{Photon Uncertainties}
The uncertainties related to overall normalization of the \gammaiso~\pt~spectra (such as luminosity scale, vertex reconstruction efficiency, trigger efficiency, and photon reconstruction efficiency) cancel completely because the observable is normalized per measured photon. Consequently, no systematic uncertainty from these sources is assigned. 

Sources of systematic uncertainty related to the photon energy scale, photon energy resolution and material budget are negligible. While the measurement is, by construction, totally insensitive to overall normalization, it is, in principle, sensitive to bin-migration or scale uncertainties that affect the shape of the photon \pt~spectra. This potential systematic uncertainty is reduced by integrating over a large photon \pt~range (12--40 \GeVc). Moreover, the EMCal performance is such that these effects are small; for a 12~GeV cluster, the resolution $\sigma/E = 1.7\% \oplus 11.3\%/\sqrt{E} \oplus 4.8\%/E$ yields $\sigma_{E}/E =3.6\%$. For a 40 GeV cluster, this yields $\sigma_{E}/E =2.4\%$.

The EMCal energy scale has been studied with test-beam data~\cite{Allen:2009aa} as well as with measurements of the energy-to-momentum ratio of electrons in $\pi^{0}\to\gamma\gamma$ events in data and simulation~\cite{Adam:2016khe}. The calorimeter uncertainty is 0.8$\%$. The uncertainties due to photon energy scale, resolution, and material budget have been estimated for the isolated photon cross section measurement with 7 TeV pp 
 and are less than 3$\%$ in the \pt~range covered in this analysis~\cite{Acharya:2019jkx}. The effects on the trigger-normalized correlation functions would be even smaller, as explained earlier in this section. Given that this level of uncertainty is much smaller than other sources of systematic uncertainties for this measurement, it is neglected.

\section{Results and Discussion}
\label{sec:results}
\begin{figure*}
     \centering
     \includegraphics[width=0.3\textwidth]{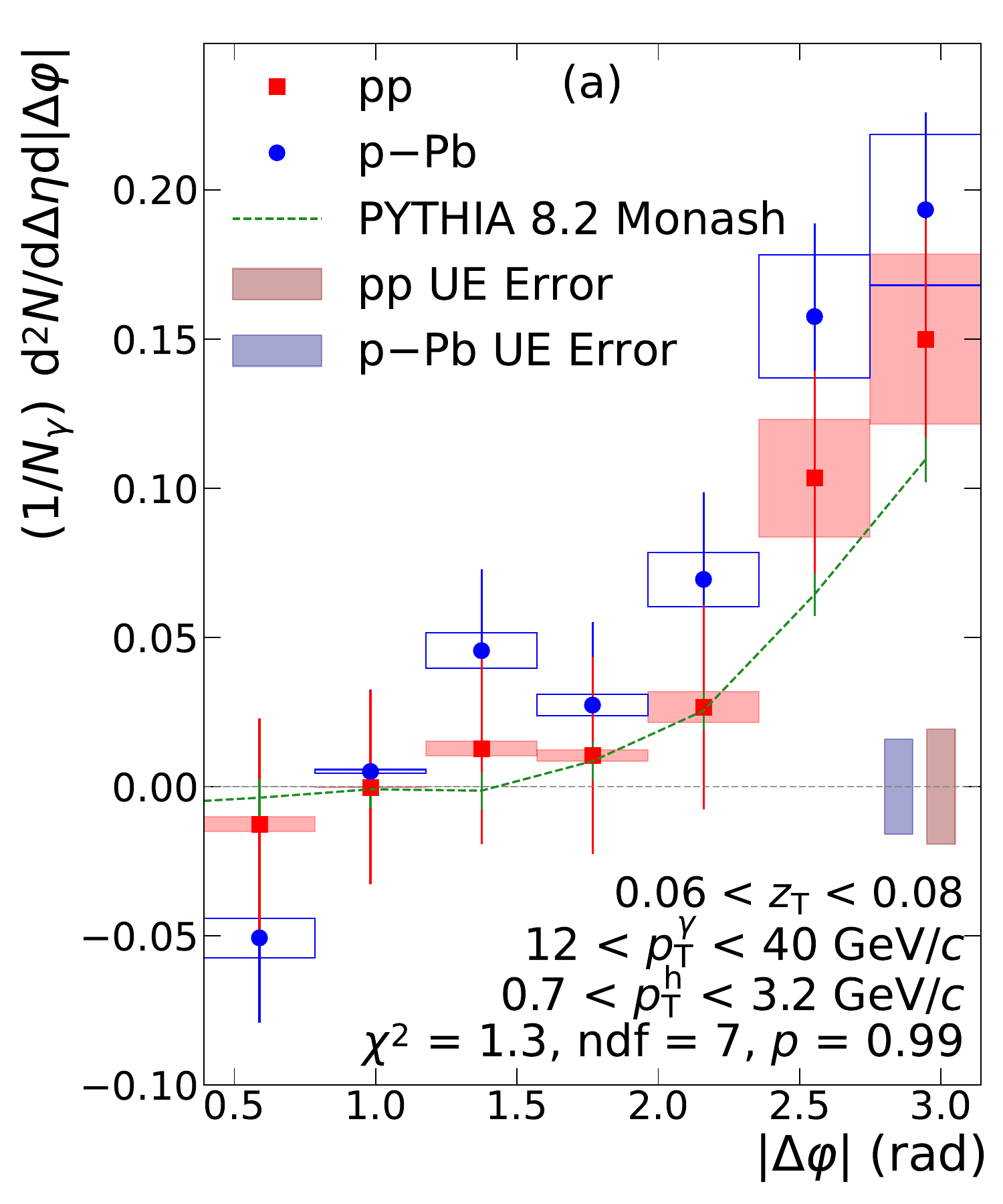}
    \includegraphics[width=0.3\textwidth]{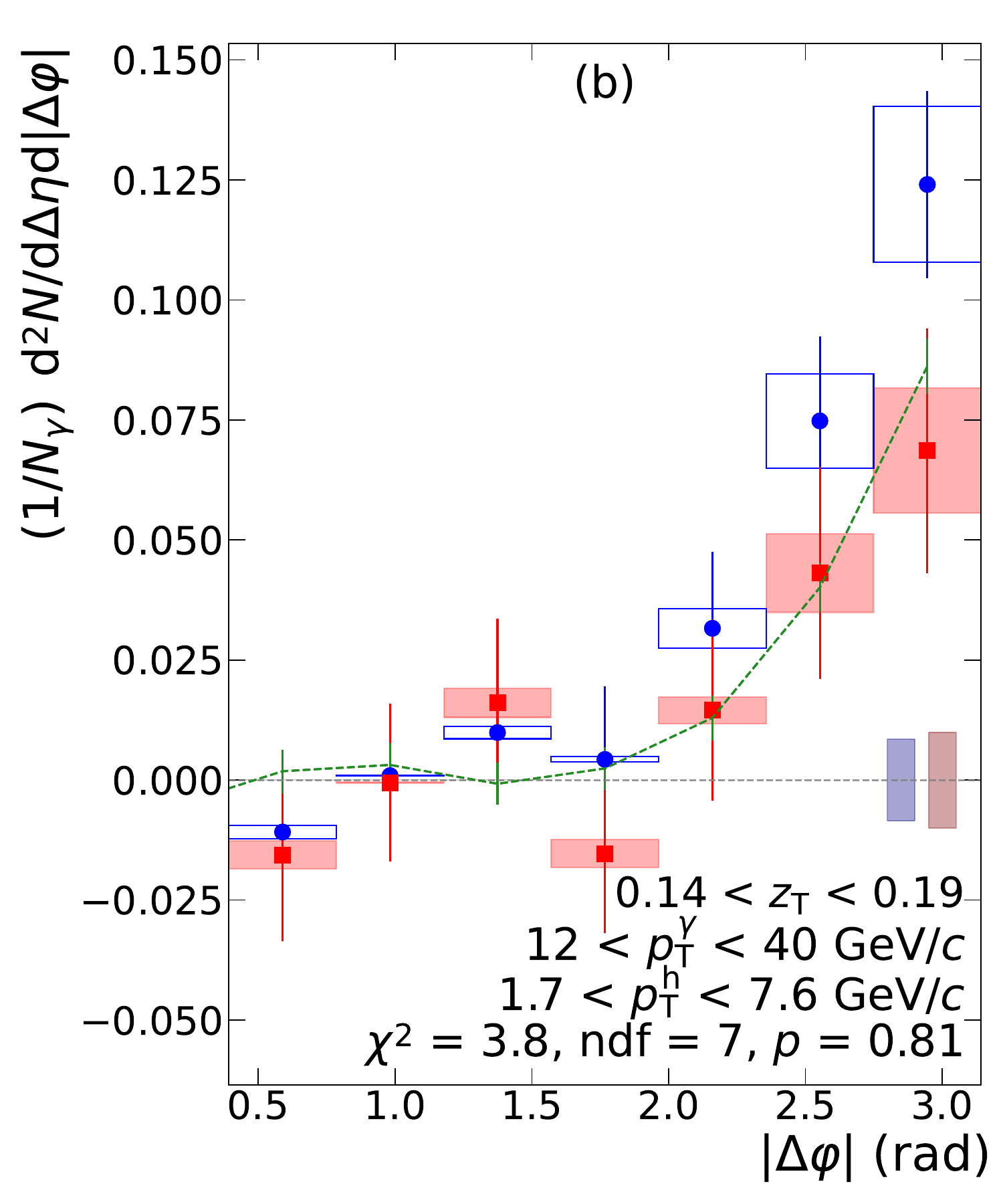}        
    \includegraphics[width=0.3\textwidth]{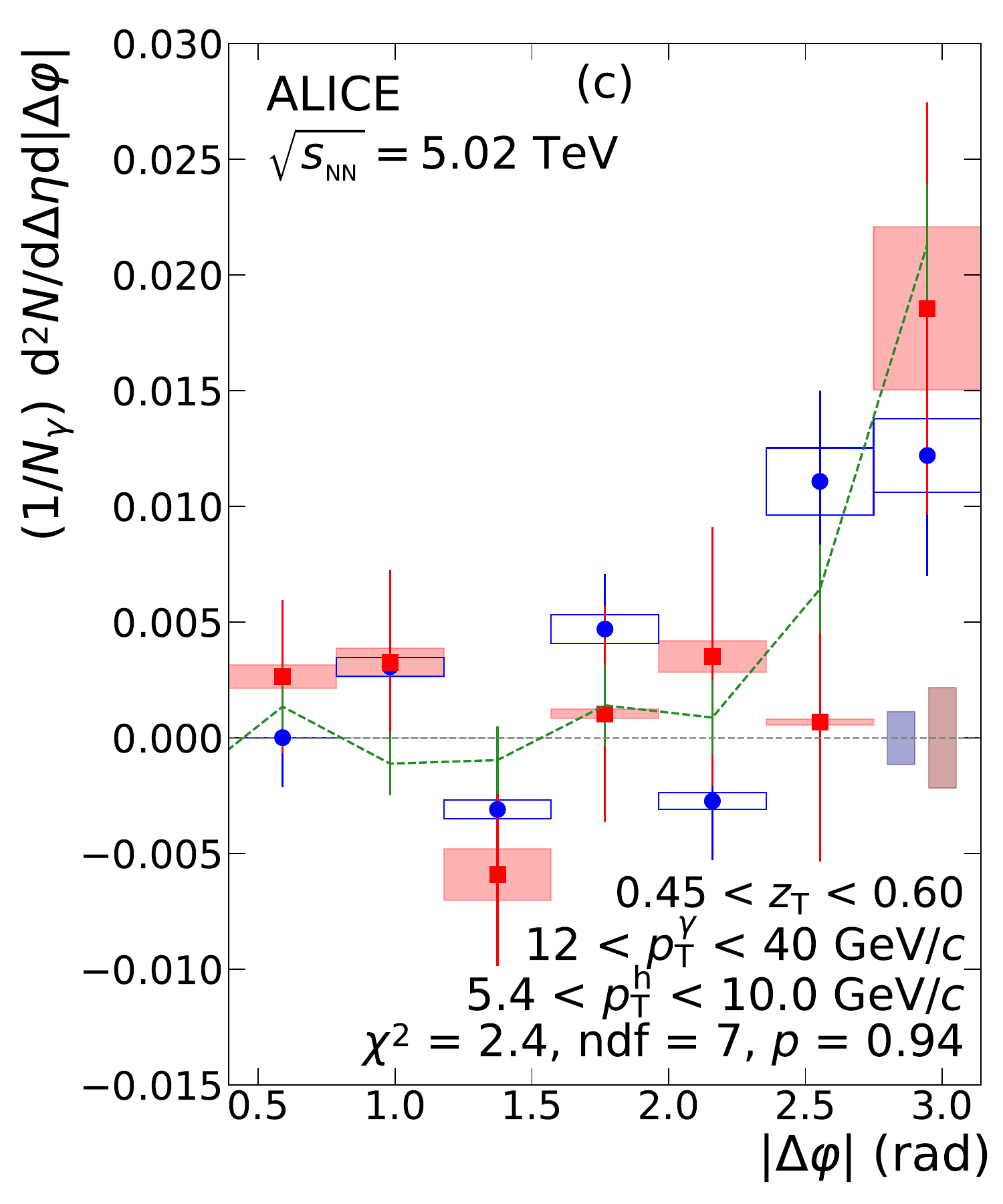}
    \caption{$\gammaiso$--hadron correlation functions for pp (red) and \pPb~(blue) data at $\sqrt{s_\mathrm{NN}}$ = 5.02 TeV as measured by the ALICE detector. The different panels represent three different \zt~bins. The correlation functions are projected over the range $|\Delta\eta| < 1.2$. The darker bands at zero represents the uncertainty from the underlying event estimation in pp and \pPb. The underlying event was estimated over the range $0.4 <|\Delta\varphi| < 1.6$. The vertical bars represent statistical uncertainties only. The boxes indicate the systematic uncertainties. The dashed green line represents the \gammaiso--hadron correlation function obtained with \textsc{PYTHIA 8.2} Monash Tune. ``$p$" is the p-value for the hypothesis that the pp and \pPb data follow the same true correlation function.
    }
     \label{fig:GH_Correlations}
 \end{figure*}

The final $ \gammaiso$-hadron correlations are reported in $\zt$  bins for each trigger-photon $\pt$ bin, where $\zt$ is the ratio of the associated hadron, $\pt^\mathrm{h}$, to isolated photon transverse momentum, $\zt = \pt^{\mathrm{h}}/\pt^{\gammaiso}$. The fully subtracted azimuthal correlations as a function of $ \Delta\varphi$, the azimuthal angle between the photon and the hadron, are shown in Fig.~\ref{fig:GH_Correlations} for pp and \pPb~data. With the measured \gammaiso~ constraining the parton kinematics, the distribution of away-side associated hadrons with momentum fraction \zt represents the fragmentation function of the parton.

\FloatBarrier


The darker colored bands at zero represents the uncertainty from the uncorrelated background estimate. The vertical bars indicate the statistical uncertainty only. The final correlation functions in each collision system demonstrate similar behavior: both show a signal consistent with zero at small $\Delta\varphi$, and a rising away-side peak at large $\Delta\varphi$ arising predominantly from the hard-scattered parton opposite to the trigger photon.

Agreement within uncertainties between pp, \pPb, and the \textsc{PYTHIA 8.2} Monash Tune is observed.
By measuring associated hadrons, correlations can be observed for much larger angles than would otherwise be possible for hadrons within a reconstructed jet. A $\chi^2$ test between pp and \pPb~data and a p-value is calculated in each \zt bin for the null hypothesis that pp and \pPb data follow the same true correlation function. In each bin, the null hypothesis cannot be rejected, indicating that there is no significant difference between the correlation functions in the two collision systems.

\begin{figure}
    \centering
    \includegraphics[width=0.67\textwidth]{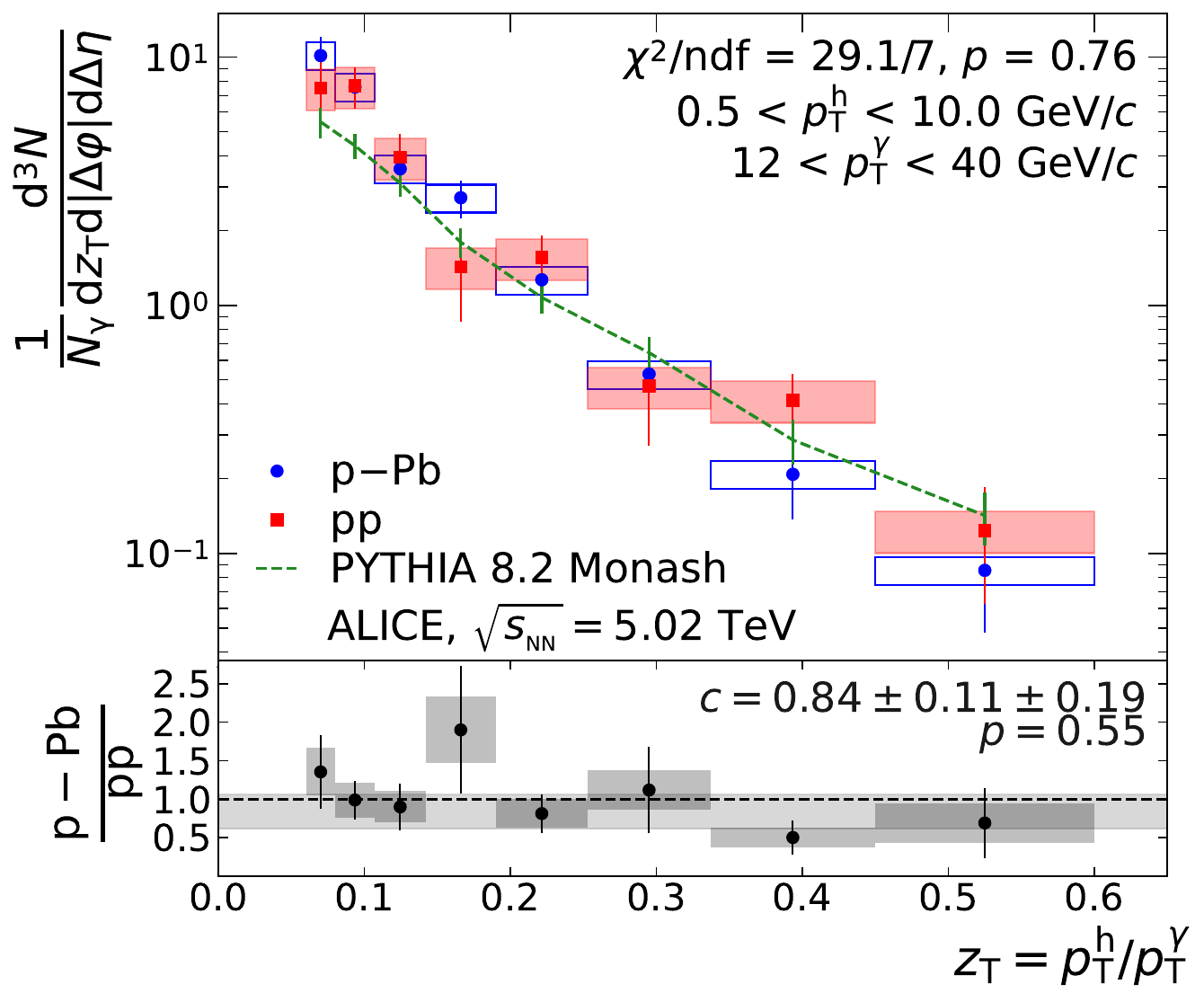}
    \caption{$\gammaiso$-tagged fragmentation function for pp (red) and \pPb~data (blue) at $\sqrt{s_\mathrm{NN}}$ = 5.02 TeV as measured by the ALICE detector. The boxes represent the systematic uncertainties while the vertical bars indicate the statistical uncertainties. The dashed green line corresponds to \textsc{PYTHIA 8.2}. The $\chi^2$ test for the comparison of pp and \pPb~data incorporates correlations among different \zt~intervals. A constant that was fit to the ratio including statistical and systematic uncertainties is shown as grey band, with the width indicating the uncertainty on the fit.}
    \label{fig:Fragmentation_Functions}
\end{figure}

The correlation functions from Fig. \ref{fig:GH_Correlations} are then integrated in the region $|\Delta\varphi| > \frac{7\pi}{8}$ for each $\zt$ bin to obtain the $\gammaiso$-tagged fragmentation function shown in Fig. \ref{fig:Fragmentation_Functions}. This range roughly corresponds to the azimuthal angle consistent with the commonly used radius of $R=$ 0.4 for jet measurements.

The statistical uncertainty on the away-side yields in each $\zt$ bin is calculated from the statistical uncertainty in the fully subtracted correlation functions, along with the statistical uncertainty arising from the uncorrelated background subtraction. A maximum charged hadron \pt of 10 \GeVc and a photon trigger \pt up to 40 \GeVc could result in a potential bias of the associated \zt spectrum. However, by repeating the analysis in different photon trigger \pt bins, it was found that any such effects were negligible compared to other uncertainties. The two largest sources of systematic uncertainty are from the purity and the single track correction factors. For the chosen $\pt^{\mathrm{track}}$ interval, there is no strong $\pt$ dependence for the uncertainty of the charged tracking efficiency.

 The ratio of the fragmentation functions in \pPb~ and pp collisions is shown in the lower panel of Fig.~\ref{fig:Fragmentation_Functions}.
The fit yields a constant factor of $0.84\pm0.11\mathrm{(stat)}\pm0.19\mathrm{(sys)}$. 
 Thus, within total uncertainties,
 the \pPb to pp ratio is consistent with unity. 


\section{Conclusions}
\label{sec:conclusions}
We report a measurement of azimuthal correlations between isolated photons and associated charged hadrons in \pPb and pp collisions at 5.02 TeV per nucleon. 
We observe no difference in the \zt~distribution between pp and \pPb data within a \zt-integrated statistical uncertainty of 13$\%$ on the ratio. 
PYTHIA 8.2 Monash Tune describes both data sets within the current precision. This measurement provides a constraint on the impact of cold nuclear matter effects on parton fragmentation, and indicates that modifications in the \zt distributions observed in Pb--Pb collisions larger than the overall uncertainty on this measurement of approximately 25\% must be due to hot medium modifications. Analysis of isolated photon-hadron correlations in Pb--Pb collisions will allow hot nuclear matter effects to be quantified. Furthermore, the next LHC run will significantly improve sensitivity to cold nuclear matter effects due to upgrades of the ALICE tracker and readout.

This measurement significantly extends previous LHC results by focusing on the fragmentation of photon-tagged low-\pt jets that probe values of {$x_{\mathrm{T}} = 2\pt/\sqrt{s_{\mathrm{NN}}}\ = $ 0.005--0.016}, which is similar to the range probed by measurements of dihadron production at forward rapidity in d--Au collisions by PHENIX that showed strong modification of the away-side yield~\cite{Adare:2011sc}. It also represents a benchmark for future measurements of jet modification in electron-nucleus scattering at the Electron-Ion Collider~\cite{Accardi:2012qut}, which will probe a similar range in Bjorken-$x$~\cite{Arratia:2019vju}.


 \FloatBarrier
\newenvironment{acknowledgement}{\relax}{\relax}
\begin{acknowledgement}
\section*{Acknowledgements}

The ALICE Collaboration would like to thank all its engineers and technicians for their invaluable contributions to the construction of the experiment and the CERN accelerator teams for the outstanding performance of the LHC complex.
The ALICE Collaboration gratefully acknowledges the resources and support provided by all Grid centres and the Worldwide LHC Computing Grid (WLCG) collaboration.
The ALICE Collaboration acknowledges the following funding agencies for their support in building and running the ALICE detector:
A. I. Alikhanyan National Science Laboratory (Yerevan Physics Institute) Foundation (ANSL), State Committee of Science and World Federation of Scientists (WFS), Armenia;
Austrian Academy of Sciences, Austrian Science Fund (FWF): [M 2467-N36] and Nationalstiftung f\"{u}r Forschung, Technologie und Entwicklung, Austria;
Ministry of Communications and High Technologies, National Nuclear Research Center, Azerbaijan;
Conselho Nacional de Desenvolvimento Cient\'{\i}fico e Tecnol\'{o}gico (CNPq), Financiadora de Estudos e Projetos (Finep), Funda\c{c}\~{a}o de Amparo \`{a} Pesquisa do Estado de S\~{a}o Paulo (FAPESP) and Universidade Federal do Rio Grande do Sul (UFRGS), Brazil;
Ministry of Education of China (MOEC) , Ministry of Science \& Technology of China (MSTC) and National Natural Science Foundation of China (NSFC), China;
Ministry of Science and Education and Croatian Science Foundation, Croatia;
Centro de Aplicaciones Tecnol\'{o}gicas y Desarrollo Nuclear (CEADEN), Cubaenerg\'{\i}a, Cuba;
Ministry of Education, Youth and Sports of the Czech Republic, Czech Republic;
The Danish Council for Independent Research | Natural Sciences, the VILLUM FONDEN and Danish National Research Foundation (DNRF), Denmark;
Helsinki Institute of Physics (HIP), Finland;
Commissariat \`{a} l'Energie Atomique (CEA) and Institut National de Physique Nucl\'{e}aire et de Physique des Particules (IN2P3) and Centre National de la Recherche Scientifique (CNRS), France;
Bundesministerium f\"{u}r Bildung und Forschung (BMBF) and GSI Helmholtzzentrum f\"{u}r Schwerionenforschung GmbH, Germany;
General Secretariat for Research and Technology, Ministry of Education, Research and Religions, Greece;
National Research, Development and Innovation Office, Hungary;
Department of Atomic Energy Government of India (DAE), Department of Science and Technology, Government of India (DST), University Grants Commission, Government of India (UGC) and Council of Scientific and Industrial Research (CSIR), India;
Indonesian Institute of Science, Indonesia;
Centro Fermi - Museo Storico della Fisica e Centro Studi e Ricerche Enrico Fermi and Istituto Nazionale di Fisica Nucleare (INFN), Italy;
Institute for Innovative Science and Technology , Nagasaki Institute of Applied Science (IIST), Japanese Ministry of Education, Culture, Sports, Science and Technology (MEXT) and Japan Society for the Promotion of Science (JSPS) KAKENHI, Japan;
Consejo Nacional de Ciencia (CONACYT) y Tecnolog\'{i}a, through Fondo de Cooperaci\'{o}n Internacional en Ciencia y Tecnolog\'{i}a (FONCICYT) and Direcci\'{o}n General de Asuntos del Personal Academico (DGAPA), Mexico;
Nederlandse Organisatie voor Wetenschappelijk Onderzoek (NWO), Netherlands;
The Research Council of Norway, Norway;
Commission on Science and Technology for Sustainable Development in the South (COMSATS), Pakistan;
Pontificia Universidad Cat\'{o}lica del Per\'{u}, Peru;
Ministry of Science and Higher Education, National Science Centre and WUT ID-UB, Poland;
Korea Institute of Science and Technology Information and National Research Foundation of Korea (NRF), Republic of Korea;
Ministry of Education and Scientific Research, Institute of Atomic Physics and Ministry of Research and Innovation and Institute of Atomic Physics, Romania;
Joint Institute for Nuclear Research (JINR), Ministry of Education and Science of the Russian Federation, National Research Centre Kurchatov Institute, Russian Science Foundation and Russian Foundation for Basic Research, Russia;
Ministry of Education, Science, Research and Sport of the Slovak Republic, Slovakia;
National Research Foundation of South Africa, South Africa;
Swedish Research Council (VR) and Knut \& Alice Wallenberg Foundation (KAW), Sweden;
European Organization for Nuclear Research, Switzerland;
Suranaree University of Technology (SUT), National Science and Technology Development Agency (NSDTA) and Office of the Higher Education Commission under NRU project of Thailand, Thailand;
Turkish Atomic Energy Agency (TAEK), Turkey;
National Academy of  Sciences of Ukraine, Ukraine;
Science and Technology Facilities Council (STFC), United Kingdom;
National Science Foundation of the United States of America (NSF) and United States Department of Energy, Office of Nuclear Physics (DOE NP), United States of America.
\end{acknowledgement}

\bibliographystyle{utphys}   
\bibliography{biblio}

\newpage
\appendix

%
%

\section{The ALICE Collaboration}
\label{app:collab}

\begingroup
\small
\begin{flushleft}
S.~Acharya\Irefn{org141}\And 
D.~Adamov\'{a}\Irefn{org95}\And 
A.~Adler\Irefn{org74}\And 
J.~Adolfsson\Irefn{org81}\And 
M.M.~Aggarwal\Irefn{org100}\And 
G.~Aglieri Rinella\Irefn{org34}\And 
M.~Agnello\Irefn{org30}\And 
N.~Agrawal\Irefn{org10}\textsuperscript{,}\Irefn{org54}\And 
Z.~Ahammed\Irefn{org141}\And 
S.~Ahmad\Irefn{org16}
% ...
\And
\renewcommand\labelenumi{\textsuperscript{\theenumi}~}

\section*{Affiliation notes}
\renewcommand\theenumi{\roman{enumi}}
\begin{Authlist}
\item \Adef{org*}Deceased
\item \Adef{orgI}Italian National Agency for New Technologies, Energy and Sustainable Economic Development (ENEA), Bologna, Italy
\item \Adef{orgII}Dipartimento DET del Politecnico di Torino, Turin, Italy
\item \Adef{orgIII}M.V. Lomonosov Moscow State University, D.V. Skobeltsyn Institute of Nuclear, Physics, Moscow, Russia
\item \Adef{orgIV}Department of Applied Physics, Aligarh Muslim University, Aligarh, India
\item \Adef{orgV}Institute of Theoretical Physics, University of Wroclaw, Poland
\end{Authlist}

\section*{Collaboration Institutes}
\renewcommand\theenumi{\arabic{enumi}~}
\begin{Authlist}
\item \Idef{org1}A.I. Alikhanyan National Science Laboratory (Yerevan Physics Institute) Foundation, Yerevan, Armenia
\item \Idef{org2}Bogolyubov Institute for Theoretical Physics, National Academy of Sciences of Ukraine, Kiev, Ukraine
\item \Idef{org3}Bose Institute, Department of Physics  and Centre for Astroparticle Physics and Space Science (CAPSS), Kolkata, India
\item \Idef{org4}Budker Institute for Nuclear Physics, Novosibirsk, Russia
\item \Idef{org5}California Polytechnic State University, San Luis Obispo, California, United States
\item \Idef{org6}Central China Normal University, Wuhan, China
\item \Idef{org7}Centre de Calcul de l'IN2P3, Villeurbanne, Lyon, France
\item \Idef{org8}Centro de Aplicaciones Tecnol\'{o}gicas y Desarrollo Nuclear (CEADEN), Havana, Cuba
\item \Idef{org9}Centro de Investigaci\'{o}n y de Estudios Avanzados (CINVESTAV), Mexico City and M\'{e}rida, Mexico
\item \Idef{org10}Centro Fermi - Museo Storico della Fisica e Centro Studi e Ricerche ``Enrico Fermi', Rome, Italy
\item \Idef{org11}Chicago State University, Chicago, Illinois, United States
\item \Idef{org12}China Institute of Atomic Energy, Beijing, China
\item \Idef{org13}Comenius University Bratislava, Faculty of Mathematics, Physics and Informatics, Bratislava, Slovakia
\item \Idef{org14}COMSATS University Islamabad, Islamabad, Pakistan
\item \Idef{org15}Creighton University, Omaha, Nebraska, United States
\item \Idef{org16}Department of Physics, Aligarh Muslim University, Aligarh, India
\item \Idef{org17}Department of Physics, Pusan National University, Pusan, Republic of Korea
\item \Idef{org18}Department of Physics, Sejong University, Seoul, Republic of Korea
\item \Idef{org19}Department of Physics, University of California, Berkeley, California, United States
\item \Idef{org20}Department of Physics, University of Oslo, Oslo, Norway
\item \Idef{org21}Department of Physics and Technology, University of Bergen, Bergen, Norway
\item \Idef{org22}Dipartimento di Fisica dell'Universit\`{a} 'La Sapienza' and Sezione INFN, Rome, Italy
\item \Idef{org23}Dipartimento di Fisica dell'Universit\`{a} and Sezione INFN, Cagliari, Italy
\item \Idef{org24}Dipartimento di Fisica dell'Universit\`{a} and Sezione INFN, Trieste, Italy
\item \Idef{org25}Dipartimento di Fisica dell'Universit\`{a} and Sezione INFN, Turin, Italy
\item \Idef{org26}Dipartimento di Fisica e Astronomia dell'Universit\`{a} and Sezione INFN, Bologna, Italy
\item \Idef{org27}Dipartimento di Fisica e Astronomia dell'Universit\`{a} and Sezione INFN, Catania, Italy
\item \Idef{org28}Dipartimento di Fisica e Astronomia dell'Universit\`{a} and Sezione INFN, Padova, Italy
\item \Idef{org29}Dipartimento di Fisica `E.R.~Caianiello' dell'Universit\`{a} and Gruppo Collegato INFN, Salerno, Italy
\item \Idef{org30}Dipartimento DISAT del Politecnico and Sezione INFN, Turin, Italy
\item \Idef{org31}Dipartimento di Scienze e Innovazione Tecnologica dell'Universit\`{a} del Piemonte Orientale and INFN Sezione di Torino, Alessandria, Italy
\item \Idef{org32}Dipartimento di Scienze MIFT, Universit\`{a} di Messina, Messina, Italy
\item \Idef{org33}Dipartimento Interateneo di Fisica `M.~Merlin' and Sezione INFN, Bari, Italy
\item \Idef{org34}European Organization for Nuclear Research (CERN), Geneva, Switzerland
\item \Idef{org35}Faculty of Electrical Engineering, Mechanical Engineering and Naval Architecture, University of Split, Split, Croatia
\item \Idef{org36}Faculty of Engineering and Science, Western Norway University of Applied Sciences, Bergen, Norway
\item \Idef{org37}Faculty of Nuclear Sciences and Physical Engineering, Czech Technical University in Prague, Prague, Czech Republic
\item \Idef{org38}Faculty of Science, P.J.~\v{S}af\'{a}rik University, Ko\v{s}ice, Slovakia
\item \Idef{org39}Frankfurt Institute for Advanced Studies, Johann Wolfgang Goethe-Universit\"{a}t Frankfurt, Frankfurt, Germany
\item \Idef{org40}Fudan University, Shanghai, China
\item \Idef{org41}Gangneung-Wonju National University, Gangneung, Republic of Korea
\item \Idef{org42}Gauhati University, Department of Physics, Guwahati, India
\item \Idef{org43}Helmholtz-Institut f\"{u}r Strahlen- und Kernphysik, Rheinische Friedrich-Wilhelms-Universit\"{a}t Bonn, Bonn, Germany
\item \Idef{org44}Helsinki Institute of Physics (HIP), Helsinki, Finland
\item \Idef{org45}High Energy Physics Group,  Universidad Aut\'{o}noma de Puebla, Puebla, Mexico
\item \Idef{org46}Hiroshima University, Hiroshima, Japan
\item \Idef{org47}Hochschule Worms, Zentrum  f\"{u}r Technologietransfer und Telekommunikation (ZTT), Worms, Germany
\item \Idef{org48}Horia Hulubei National Institute of Physics and Nuclear Engineering, Bucharest, Romania
\item \Idef{org49}Indian Institute of Technology Bombay (IIT), Mumbai, India
\item \Idef{org50}Indian Institute of Technology Indore, Indore, India
\item \Idef{org51}Indonesian Institute of Sciences, Jakarta, Indonesia
\item \Idef{org52}INFN, Laboratori Nazionali di Frascati, Frascati, Italy
\item \Idef{org53}INFN, Sezione di Bari, Bari, Italy
\item \Idef{org54}INFN, Sezione di Bologna, Bologna, Italy
\item \Idef{org55}INFN, Sezione di Cagliari, Cagliari, Italy
\item \Idef{org56}INFN, Sezione di Catania, Catania, Italy
\item \Idef{org57}INFN, Sezione di Padova, Padova, Italy
\item \Idef{org58}INFN, Sezione di Roma, Rome, Italy
\item \Idef{org59}INFN, Sezione di Torino, Turin, Italy
\item \Idef{org60}INFN, Sezione di Trieste, Trieste, Italy
\item \Idef{org61}Inha University, Incheon, Republic of Korea
\item \Idef{org62}Institute for Nuclear Research, Academy of Sciences, Moscow, Russia
\item \Idef{org63}Institute for Subatomic Physics, Utrecht University/Nikhef, Utrecht, Netherlands
\item \Idef{org64}Institute of Experimental Physics, Slovak Academy of Sciences, Ko\v{s}ice, Slovakia
\item \Idef{org65}Institute of Physics, Homi Bhabha National Institute, Bhubaneswar, India
\item \Idef{org66}Institute of Physics of the Czech Academy of Sciences, Prague, Czech Republic
\item \Idef{org67}Institute of Space Science (ISS), Bucharest, Romania
\item \Idef{org68}Institut f\"{u}r Kernphysik, Johann Wolfgang Goethe-Universit\"{a}t Frankfurt, Frankfurt, Germany
\item \Idef{org69}Instituto de Ciencias Nucleares, Universidad Nacional Aut\'{o}noma de M\'{e}xico, Mexico City, Mexico
\item \Idef{org70}Instituto de F\'{i}sica, Universidade Federal do Rio Grande do Sul (UFRGS), Porto Alegre, Brazil
\item \Idef{org71}Instituto de F\'{\i}sica, Universidad Nacional Aut\'{o}noma de M\'{e}xico, Mexico City, Mexico
\item \Idef{org72}iThemba LABS, National Research Foundation, Somerset West, South Africa
\item \Idef{org73}Jeonbuk National University, Jeonju, Republic of Korea
\item \Idef{org74}Johann-Wolfgang-Goethe Universit\"{a}t Frankfurt Institut f\"{u}r Informatik, Fachbereich Informatik und Mathematik, Frankfurt, Germany
\item \Idef{org75}Joint Institute for Nuclear Research (JINR), Dubna, Russia
\item \Idef{org76}Korea Institute of Science and Technology Information, Daejeon, Republic of Korea
\item \Idef{org77}KTO Karatay University, Konya, Turkey
\item \Idef{org78}Laboratoire de Physique des 2 Infinis, Ir\`{e}ne Joliot-Curie, Orsay, France
\item \Idef{org79}Laboratoire de Physique Subatomique et de Cosmologie, Universit\'{e} Grenoble-Alpes, CNRS-IN2P3, Grenoble, France
\item \Idef{org80}Lawrence Berkeley National Laboratory, Berkeley, California, United States
\item \Idef{org81}Lund University Department of Physics, Division of Particle Physics, Lund, Sweden
\item \Idef{org82}Nagasaki Institute of Applied Science, Nagasaki, Japan
\item \Idef{org83}Nara Women{'}s University (NWU), Nara, Japan
\item \Idef{org84}National and Kapodistrian University of Athens, School of Science, Department of Physics , Athens, Greece
\item \Idef{org85}National Centre for Nuclear Research, Warsaw, Poland
\item \Idef{org86}National Institute of Science Education and Research, Homi Bhabha National Institute, Jatni, India
\item \Idef{org87}National Nuclear Research Center, Baku, Azerbaijan
\item \Idef{org88}National Research Centre Kurchatov Institute, Moscow, Russia
\item \Idef{org89}Niels Bohr Institute, University of Copenhagen, Copenhagen, Denmark
\item \Idef{org90}Nikhef, National institute for subatomic physics, Amsterdam, Netherlands
\item \Idef{org91}NRC Kurchatov Institute IHEP, Protvino, Russia
\item \Idef{org92}NRC \guillemotleft Kurchatov\guillemotright~Institute - ITEP, Moscow, Russia
\item \Idef{org93}NRNU Moscow Engineering Physics Institute, Moscow, Russia
\item \Idef{org94}Nuclear Physics Group, STFC Daresbury Laboratory, Daresbury, United Kingdom
\item \Idef{org95}Nuclear Physics Institute of the Czech Academy of Sciences, \v{R}e\v{z} u Prahy, Czech Republic
\item \Idef{org96}Oak Ridge National Laboratory, Oak Ridge, Tennessee, United States
\item \Idef{org97}Ohio State University, Columbus, Ohio, United States
\item \Idef{org98}Petersburg Nuclear Physics Institute, Gatchina, Russia
\item \Idef{org99}Physics department, Faculty of science, University of Zagreb, Zagreb, Croatia
\item \Idef{org100}Physics Department, Panjab University, Chandigarh, India
\item \Idef{org101}Physics Department, University of Jammu, Jammu, India
\item \Idef{org102}Physics Department, University of Rajasthan, Jaipur, India
\item \Idef{org103}Physikalisches Institut, Eberhard-Karls-Universit\"{a}t T\"{u}bingen, T\"{u}bingen, Germany
\item \Idef{org104}Physikalisches Institut, Ruprecht-Karls-Universit\"{a}t Heidelberg, Heidelberg, Germany
\item \Idef{org105}Physik Department, Technische Universit\"{a}t M\"{u}nchen, Munich, Germany
\item \Idef{org106}Politecnico di Bari, Bari, Italy
\item \Idef{org107}Research Division and ExtreMe Matter Institute EMMI, GSI Helmholtzzentrum f\"ur Schwerionenforschung GmbH, Darmstadt, Germany
\item \Idef{org108}Rudjer Bo\v{s}kovi\'{c} Institute, Zagreb, Croatia
\item \Idef{org109}Russian Federal Nuclear Center (VNIIEF), Sarov, Russia
\item \Idef{org110}Saha Institute of Nuclear Physics, Homi Bhabha National Institute, Kolkata, India
\item \Idef{org111}School of Physics and Astronomy, University of Birmingham, Birmingham, United Kingdom
\item \Idef{org112}Secci\'{o}n F\'{\i}sica, Departamento de Ciencias, Pontificia Universidad Cat\'{o}lica del Per\'{u}, Lima, Peru
\item \Idef{org113}St. Petersburg State University, St. Petersburg, Russia
\item \Idef{org114}Stefan Meyer Institut f\"{u}r Subatomare Physik (SMI), Vienna, Austria
\item \Idef{org115}SUBATECH, IMT Atlantique, Universit\'{e} de Nantes, CNRS-IN2P3, Nantes, France
\item \Idef{org116}Suranaree University of Technology, Nakhon Ratchasima, Thailand
\item \Idef{org117}Technical University of Ko\v{s}ice, Ko\v{s}ice, Slovakia
\item \Idef{org118}The Henryk Niewodniczanski Institute of Nuclear Physics, Polish Academy of Sciences, Cracow, Poland
\item \Idef{org119}The University of Texas at Austin, Austin, Texas, United States
\item \Idef{org120}Universidad Aut\'{o}noma de Sinaloa, Culiac\'{a}n, Mexico
\item \Idef{org121}Universidade de S\~{a}o Paulo (USP), S\~{a}o Paulo, Brazil
\item \Idef{org122}Universidade Estadual de Campinas (UNICAMP), Campinas, Brazil
\item \Idef{org123}Universidade Federal do ABC, Santo Andre, Brazil
\item \Idef{org124}University of Cape Town, Cape Town, South Africa
\item \Idef{org125}University of Houston, Houston, Texas, United States
\item \Idef{org126}University of Jyv\"{a}skyl\"{a}, Jyv\"{a}skyl\"{a}, Finland
\item \Idef{org127}University of Liverpool, Liverpool, United Kingdom
\item \Idef{org128}University of Science and Technology of China, Hefei, China
\item \Idef{org129}University of South-Eastern Norway, Tonsberg, Norway
\item \Idef{org130}University of Tennessee, Knoxville, Tennessee, United States
\item \Idef{org131}University of the Witwatersrand, Johannesburg, South Africa
\item \Idef{org132}University of Tokyo, Tokyo, Japan
\item \Idef{org133}University of Tsukuba, Tsukuba, Japan
\item \Idef{org134}Universit\'{e} Clermont Auvergne, CNRS/IN2P3, LPC, Clermont-Ferrand, France
\item \Idef{org135}Universit\'{e} de Lyon, Universit\'{e} Lyon 1, CNRS/IN2P3, IPN-Lyon, Villeurbanne, Lyon, France
\item \Idef{org136}Universit\'{e} de Strasbourg, CNRS, IPHC UMR 7178, F-67000 Strasbourg, France, Strasbourg, France
\item \Idef{org137}Universit\'{e} Paris-Saclay Centre d'Etudes de Saclay (CEA), IRFU, D\'{e}partment de Physique Nucl\'{e}aire (DPhN), Saclay, France
\item \Idef{org138}Universit\`{a} degli Studi di Foggia, Foggia, Italy
\item \Idef{org139}Universit\`{a} degli Studi di Pavia, Pavia, Italy
\item \Idef{org140}Universit\`{a} di Brescia, Brescia, Italy
\item \Idef{org141}Variable Energy Cyclotron Centre, Homi Bhabha National Institute, Kolkata, India
\item \Idef{org142}Warsaw University of Technology, Warsaw, Poland
\item \Idef{org143}Wayne State University, Detroit, Michigan, United States
\item \Idef{org144}Westf\"{a}lische Wilhelms-Universit\"{a}t M\"{u}nster, Institut f\"{u}r Kernphysik, M\"{u}nster, Germany
\item \Idef{org145}Wigner Research Centre for Physics, Budapest, Hungary
\item \Idef{org146}Yale University, New Haven, Connecticut, United States
\item \Idef{org147}Yonsei University, Seoul, Republic of Korea
\end{Authlist}
\endgroup
  
\end{document}